\documentclass[10pt,journal]{IEEEtran}
\IEEEoverridecommandlockouts

\usepackage{graphicx,subcaption}
\usepackage{amsmath,amsthm,amsfonts,amssymb}
\usepackage{cite}
\usepackage{bm}
\usepackage{bbm}
\usepackage{url}
\usepackage{array}
\usepackage{color}
\usepackage{multirow}
\usepackage{booktabs}
\usepackage[table,xcdraw]{xcolor}
\usepackage{enumitem}

\usepackage{makecell}

\theoremstyle{plain}
\newtheorem{thm}{Theorem}
\newtheorem{lem}[thm]{Lemma}

\newtheorem{prop}[thm]{Proposition}

\newtheorem{sty1}{Theorem}
\newtheorem{defi}[sty1]{Definition}

\newenvironment{NewProof}{{\noindent\it Proof.}}{\hfill $\blacksquare$\par}

\usepackage[english]{babel}
\usepackage{algorithm}
\usepackage[noend]{algpseudocode}

\allowdisplaybreaks[4]

\begin{document}
\title{Dynamic gNodeB Sleep Control for Energy-Conserving 5G Radio Access Network}

\author{
Pengfei~Shen,
Yulin~Shao,~\IEEEmembership{Member,~IEEE},
Qi~Cao,
Lu Lu,~\IEEEmembership{Member,~IEEE}

\thanks{P. Shen and L. Lu are with the University of Chinese Academy of Sciences, and the Key Laboratory of Space Utilization, Chinese Academy of Sciences, Beijing 100094, China (emails: \{shenpengfei19, lulu\}@csu.ac.cn).

Y. Shao is with the Department of Electrical and Electronic Engineering, Imperial College London, London SW7 2AZ, U.K. (e-mail: y.shao@imperial.ac.uk). 

C. Qi is with Xidian-Guangzhou Research Institute, Xidian University, Guangzhou, China (e-mail: caoqi@xidian.edu.cn).
}
}

\maketitle
\begin{abstract}
5G radio access network (RAN) is consuming much more energy than legacy RAN due to the denser deployments of gNodeBs (gNBs) and higher single-gNB power consumption.
In an effort to achieve an energy-conserving RAN, this paper develops a dynamic on-off switching paradigm, where the ON/OFF states of gNBs can be dynamically configured according to the evolvements of the associated users.
We formulate the dynamic sleep control for a cluster of gNBs as a Markov decision process (MDP) and analyze various switching policies to reduce the energy expenditure.
The optimal policy of the MDP that minimizes the energy expenditure can be derived from dynamic programming, but the computation is expensive.
To circumvent this issue, this paper puts forth a greedy policy and an index policy for gNB sleep control.
When there is no constraint on the number of gNBs that can be turned off, we prove the dual-threshold structure of the greedy policy and analyze its connections with the optimal policy.
Inspired by the dual-threshold structure and Whittle index, we develop an index policy by decoupling the original MDP into multiple one-dimensional MDPs -- the indexability of the decoupled MDP is proven and an algorithm to compute the index is proposed.
Extensive simulation results verify that the index policy exhibits close-to-optimal performance in terms of the energy expenditure of the gNB cluster.
As far as the computational complexity is concerned, on the other hand, the index policy is much more efficient than the optimal policy, which is computationally prohibitive when the number of gNBs is large.
\end{abstract}


\section{Introduction}\label{sec:I}
\subsection{Background}
With the rolling out of 5G new radio (NR), the energy expenditure of commercial broadband cellular networks becomes a growing concern \cite{en14175392,9678321,ISRAR2021102910,FLOAC}.
To support enhanced mobile broadband communications at $10$ Gbps and provide seamless coverage, 5G base stations (BSs), i.e., gNodeB (gNB), are deployed more densely than 4G eNodeB (eNB).
It is anticipated that the number of gNB will reach 65 million by 2025, and the average density of gNB will be three times higher than that of eNB \cite{han2020energy}.
On the other hand, gNB incorporates a number of new and power-hungry components \cite{3gpp.21.915}, such as integrated massive MIMO antennas, faster data converters, high-power/low-noise amplifiers, and millimeter wave (mmWave) transceivers.
As a consequence, the power expenditure of a single gNB is estimated to be two to four times higher than that of an eNB \cite{HUawei5GPower}. The increasing energy expenditure of mobile radio access networks (RANs) leads to higher costs and larger amounts of greenhouse gas emissions -- as it stands now, the mobile communication industry contributes 15\% to 20\% of CO$_2$ emissions among the information and communication technology (ICT) industries \cite{6512843}.

To reduce the overall cost and achieve green mobile networks, various energy-conserving schemes have been proposed in the literature  \cite{7446253,SALAHDINE2021108567,PNC,IFDMA}, such as BS sleep control, energy harvesting, hardware optimization, ene\-rgy-effi\-ciency-ori\-ented resource allocation, and network coverage planning, among which BS sleep control receives the most attention.
The power consumption of a BS mainly comes from three aspects:
1) the static consumption \cite{5992823,7445140,deruyck2014power}, i.e., the power consumed by operating the BS, such as the power consumption of the electrical parts, circuits, cooling systems, etc.;
2) the dynamic consumption \cite{5992823,7445140,deruyck2014power,9606296,woon2021peak}, i.e., the power consumed by serving users;
and 3) the switching consumption \cite{9340607,8475536,gong2012dynamic}, i.e., the power consumed by switching the BSs between ON and OFF states.
The traffic load of a BS can vary dramatically at different times of day \cite{7445140,9678321}. The deployment and operation of BSs, however, are often designed to meet the peak traffic load -- hence a large amount of energy for operating the BSs is wasted during off-peak hours.
In this light, BS sleep control is proposed to turn off the BSs with no or light traffic and wake them up again when there is moderate or heavy traffic, thereby saving the static power of light-traffic BSs.
A practice of BS sleep control is implemented by China Mobile \cite{8014292}, in which a fraction of BSs are manually turned off overnight. It is reported that the energy expenditure is reduced by 36 million kilowatt hour (kWh) per year with the manual configuration of BS states.

\subsection{Prior arts}
Many research efforts have been devoted to BS sleep control in heterogeneous mobile networks \cite{7446253,7539375}. 
In general, the problem of BS switching control is combinatorial optimization and is often NP-hard.
The research focus of prior arts is on designing efficient BS switching policies.

Early studies on BS sleep control utilize state-independent policies \cite{6774407,7060678}, such as random policy, to control the sleep mode of either macro or micro BSs in heterogeneous networks.
In their formulations, the locations of mobile users, macro and micro BSs are often modeled as independent homogeneous Poisson point processes (PPPs). Each BS is associated with a turn-off probability.
The optimization problems are then formulated to find the optimal set of turn-off probabilities to minimize the BS energy expenditure.

Beyond state-independent policies, a more prevailing approach to model the problem of BS sleep control is taking user distribution, BS traffic load, or transmission power budget, etc., as ``states'' and designing switching policies that are governed by these states.
Under this formulation, \cite{9435313,Yu2014,8866747} investigated threshold-based switching policies; \cite{5992823,6786060,6996018} investigated the greedy policy; \cite{6489498}, \cite{7543472}, and \cite{9764674} developed a switching-on/off based energy saving (SWES) algorithm, a local search algorithm, and a genetic algorithm (GA)-based algorithm, respectively, to solve the combinatorial optimization problems.
These works, however, aim to minimize the myopic energy expenditure of mobile networks, hence is suboptimal in terms of the long-term energy expenditure when the network dynamics are correlated over time.

To develop switching policies that minimize the long-term network energy expenditure, \cite{8935062,6907934,8777099,9340607,8735834,6747280,9209916} formulate the BS sleep control problem as Markov decision processes (MDPs).
Specifically, the optimal policy for the MDP is investigated in \cite{8935062} and \cite{6907934}. The optimal policy, however, is known for its high complexity. To circumvent this issue, \cite{8777099} proposed a policy rollout algorithm to be used in conjunction with the greedy policy to approximate the value function of each state.
Along this direction, another line of work leverages deep reinforcement learning (DRL) techniques to solve the MDP \cite{9340607,8735834,6747280,9209916}. The salient feature of DRL is model-free. That is, given unknown network dynamics, DRL algorithms are able to adapt to the traffic load variations and produce good switching policies that reduce the network power consumption \cite{puterman2014markov,Significant2020}. DRL algorithms, however, only produce achievability under an unknown environment, but fail to characterize the optimal policy of MDP.

\subsection{Contributions}
\begin{figure}[t]
\centering
\includegraphics[width=1\columnwidth]{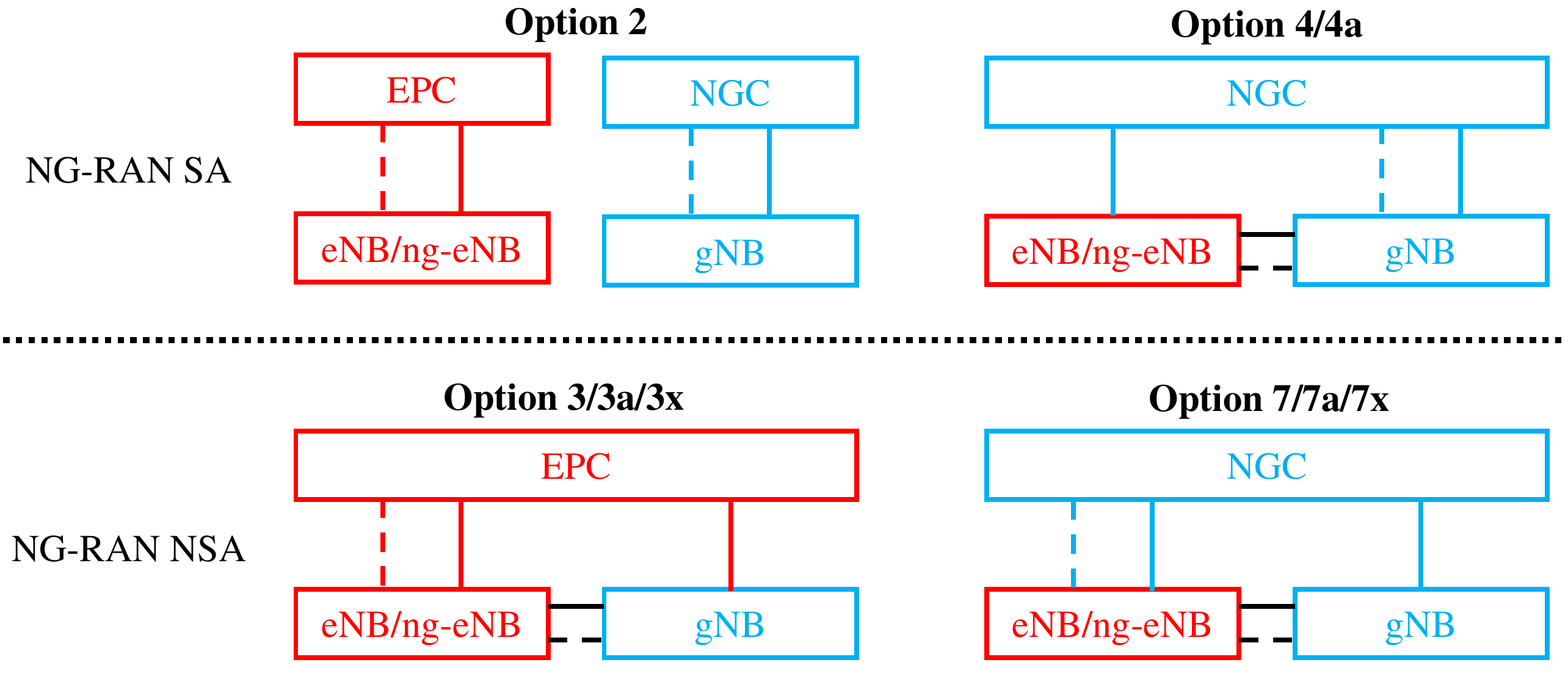}
\caption{The standalone (SA) and non-standalone (NSA) architectures of NG-RAN. The difference between SA and NSA lies in whether the control plane of gNB is connected to the core networks directly or via eNB/ng-eNB. The core network can be 4G evolved packet core (EPC) or 5G new generation core (NGC).}
\label{fig:network_options}
\end{figure}

In this paper, we put forth a dynamic on-off switching approach to achieve efficient BS sleep control tailored for the new generation RAN (NG-RAN) architecture standardized in the latest third generation partnership project (3GPP) releases.
As shown in Fig.~\ref{fig:network_options}, 3GPP NG-RAN defines two classes of RAN architectures for the deployment of 5G NR \cite{3gpp.38.801}, i.e., the standalone (SA) and non-standalone (NSA) architectures.
In both architectures, a user can connect to either gNB or eNB/ng-eNB (ng-eNB is an updated version of 4G eNB), indicating that light-traffic gNB can be turned off and the traffic demands of users in a turned-off gNB can be fulfilled by eNB or ng-eNB.

To achieve judicious gNB sleep control and energy-conserving 5G networks, we take the dynamic evolvement of users into account and formulate the on-off configuration of a cluster of 5G cells as an MDP.
Given this formulation, we design and analyze various switching policies to minimize the long-term average {\it cost} of the 5G cells, where the cost of a 5G cell is a non-decreasing function of the cell's power consumption.
The optimal policy to the formulated MDP exhibits the best performance, i.e., the minimum long-term average cost, but its computational complexity increases exponentially in the number of 5G cells.
State-independent policies, in contrast, are computationally simple, but their performances are suboptimal.
In this context, we propose and analyze two policies, one is a greedy policy and the other is an index policy to solve the MDP.
For the greedy policy, we prove its dual-threshold structure when there is no constraint on the number of gNBs that can be turned off, and analyze its connections with the optimal policy. 
Furthermore, inspired by the dual-threshold structure and Whittle index, we decouple the original MDP to multiple one-dimensional MDPs and prove the indexability of the decoupled MDP, whereby an index policy is proposed.
As far as the long-term average cost is concerned, the index policy achieves close-to-optimal performance in various simulation setups and is better than the greedy policy and state-independent policies.
As far as the computational complexity is concerned, the index policy is much more efficient to compute than the optimal policy and the greedy policy.

The remainder of this paper is organized as follows. Section~\ref{sec:II} presents the system model and formulates the MDP.
Section~\ref{sec:III} analyzes the optimal policy and the greedy policy.
Section~\ref{sec:IV} studies the index policy.
Two state-independent policies are analyzed in Section~\ref{sec:V}.
In addition, a lower bound is derived to measure the performance of various policies when the optimal policy is computationally prohibitive.
Numerical and simulation results are presented in Section~\ref{sec:VI}.
Section~\ref{sec:Conclusion} concludes this paper.

\textbf{Notations} -- 
We use boldface lowercase letters to denote column vectors and boldface uppercase letters to denote matrices.
For a vector or matrix, $(\cdot)^\top$ denotes the transpose.
$\mathbb{R}$ and $\mathbb{N}$ stand for the sets of real and non-negative integer values, respectively.
The imaginary unit is represented by $j$.
The cardinality of a set $\mathcal{V}$ is denoted by $|\mathcal{V}|$.

\section{System Model}\label{sec:II}
\subsection{Problem formulation}\label{sec:IIA}
We consider a cluster of $M$ 5G cells indexed by $\{m:m = 1,2, \cdots ,M\}$. Each cell is equipped with a gNB located in the center and the gNBs are managed by a central controller. The power consumption of a gNB consists of three main parts {\cite{7445140,9678321}}: static, dynamic, and switching, denoted by ${\mathcal{P}}_\text{static}$, ${\mathcal{P}}_\text{dynamic}$, and ${\mathcal{P}}_\text{switch}$, respectively. ${\mathcal{P}}_\text{static}$ is the power consumption incurred by the operations of the gNB, such as the electrical parts, circuits, cooling systems, etc. It is constant when the gNB is on and zero when the gNB is off. ${\mathcal{P}}_\text{dynamic}$ is the power consumed by serving users, and hence, is proportional to the number of users in the cell. ${\mathcal{P}}_\text{switch}$ is the power consumed by state switching when the gNB is turned from the OFF state to the ON state (the power consumption of switching state from ON to OFF is often omitted \cite{9340607}).


\begin{figure}[t]
\centering
\includegraphics[width=1\columnwidth]{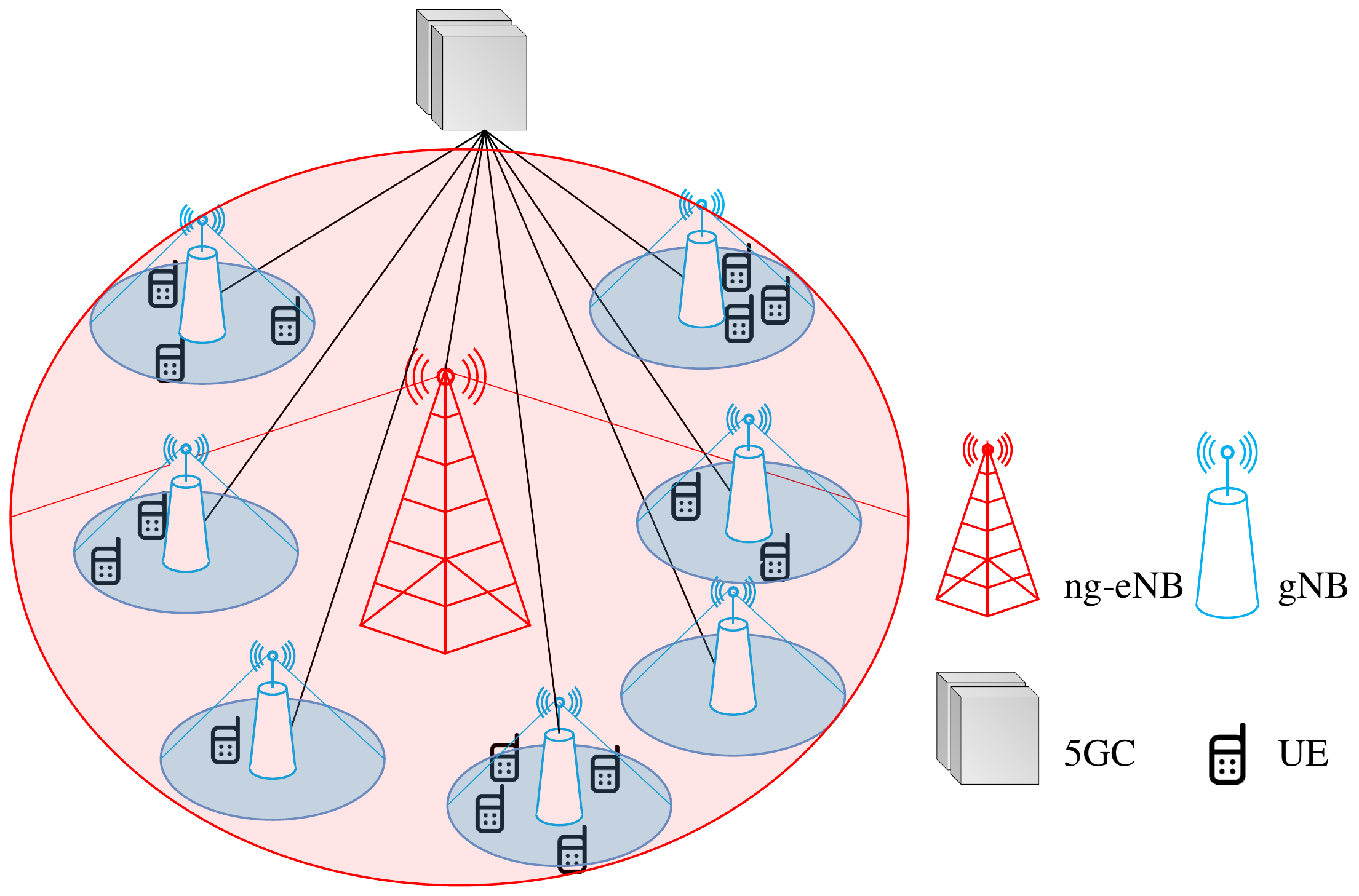}
\caption{A cluster of 5G cells with $M$ gNBs and one ng-eNB, where NGC stands for the 5G core network and UE stands for user.}
\label{fig:system_model}
\end{figure}

Time is divided into segments of duration $T_s$. At the beginning of a time segment, the central controller configures the ON/OFF states of the gNBs based on the number of existing users in the cells as well as the anticipated users in the future. Intuitively, when the number of users in a cell is large, the gNB has to be turned on to provide good quality of services to users at the expense of high power expenditure. On the other hand, when the number of users is small, the gNB can be turned off to save ${\mathcal{P}}_\text{static}$ and ${\mathcal{P}}_\text{dynamic}$. In this case, the users in the 5G cells will be served by the ng-eNB, which is capable of satisfying most requirements of the small number of users. Note that ng-eNB is always on, thus, there is no switching power consumption and its static power consumption is irrelevant. The dynamic power consumption of the ng-eNB is denoted by ${\mathcal{P}}_\text{extra}$, which is proportional to the number of users to be served (as ${\mathcal{P}}_\text{dynamic}$). Typically, the per-user dynamic power consumption of a gNB is smaller than that of the ng-eNB thanks to the closer deployments to the users. In this paper, we assume that the ng-eNB can serve at most $K$ 5G cells and $0\le K\le M$. Note that when $K = M$, all $M$ gNBs are allowed to be turned off.

To discover the optimal configuration policy for the central controller, this paper models the state configuration problem of the cell cluster as a discrete MDP, wherein the central controller makes successive switching actions based on the states of the BS cells to minimize the long-term average power consumption. More rigorously, we define the ingredients of the MDP as follows.

\begin{defi}[Action]\label{defi_action}
At the beginning of the $t$-th time segment, the action of the central controller is defined as $\bm{a}^t \triangleq (a_1^t,a_2^t, \cdots ,a_M^t)^\top$, where $a_m^t \in \{0,1\}$ denotes the state of the $m$-th gNB in the $t$-th time segment (0 and 1 stand for OFF and ON, respectively). A constraint on $\bm{a}^t$ is $\sum_{m = 1}^M a_m^t \ge M - K$, $\forall t$, as the ng-eNB can serve at most $K$ cells, where $K \in \{0,1,2, \cdots ,M\}$.
\end{defi}

\begin{defi}[State]\label{defi_state}
At the beginning of the $t$-th time segment, the state of the cell cluster is defined as $\bm{s}^t  \triangleq (s_1^t,s_2^t, \cdots ,s_M^t)^\top$, where $s_m^t$ denotes the state of the $m$-th cell. In particular, $s_m^t \triangleq (a_m^{t - 1},\widetilde n_m^t)$ consists of two parts: $a_m^{t-1}$ is the ON/OFF state of the $m$-th gNB in the $(t-1)$-th time segment and $\widetilde n_m^t$ denotes the number of users in the $m$-th cell.
\end{defi}

It is worth noting that $\widetilde n_m^t$ is the residual users from the $(t-1)$-th time segment. On the other hand, the dynamic power consumption ${\mathcal{P}}_\text{dynamic}$ is proportional to the total number of users $N_m^t$ in the cell. In the $t$-th time segment, $N_m^t$ consists of two parts: the residual users from the $(t-1)$-th time segment and the newly arrived users. Thus, we can write $N_m^t=\widetilde n_m^t+n_m^t$, where $n_m^t$ denotes the number of newly arrived users in the $t$-th time segment. The user arrival and departure models are explained in more detail later in Section~\ref{sec:IIB}.

\begin{defi}[Immediate cost]\label{defi_immediate_cost}
Suppose the power consumed by the $m$-th cell is 
${\mathcal{P}} (s_m^t,a_m^t)$ in the $t$-th time segment. The immediate cost of the action $a_m^t$ when in state $s_m^t$ is defined as $c(s_m^t,a_m^t) \triangleq f [{\mathcal{P}} (s_m^t,a_m^t) ]$, where $f$ is a non-decreasing function. Accordingly, the immediate cost of the cell cluster is $C(\bm{s}^t ,\bm{a}^t) \triangleq \sum_{m=1}^M c(s_m^t,a_m^t)$.
\end{defi}

A configuration/switching policy $\pi$ is a mapping from the state space to the action space, i.e., $\bm{a}^t=\pi(\bm{s}^t )$. Given the above definitions, the average cost incurred by a given policy $\pi$ over the infinite-time horizon is
\begin{equation}\label{eqA1}
    {{\overline C}_\pi } = \lim \limits_{T \to \infty } { \mathbb{E}\left[ \frac{1}{T} \sum_{t = 0}^{T - 1} C(\bm{s}^t,\bm{a}^t) \right] },
\end{equation}
where the expectation is taken over the dynamics of user arrival and departure in each cell.

The objective of the central controller is to discover the optimal policy $\pi^*$ such that the long-term average cost ${\overline{C}}_\pi$ is minimized, giving
\begin{eqnarray}\label{eqA2}
    (\text{P1}) \hspace{-0.3cm}&:&\hspace{-0.2cm} {\pi ^ * } = \arg \min \limits_\pi  {\overline{C}_\pi },\\
    s.t. \hspace{-0.3cm}&,&\hspace{-0.2cm} \sum_{m = 1}^M a_m^t \ge M-K,~\forall t. \nonumber
\end{eqnarray}

\subsection{Users' arrival, departure, and power consumption}\label{sec:IIB}
This paper models the users’ arrival and departure as follows.

\begin{defi}[User arrival and departure processes]\label{defi_user_process}
We assume that users arrive at and depart from a cell in an independent and identically distributed (i.i.d.) fashion. In particular,
\begin{enumerate}
    \item Users’ arrival to a cell follows a mixed Poisson process with a set of parameters $\Lambda=( \lambda_1, \lambda_2,\cdots,\lambda_J)$. That is, the number of newly arrived users in the t-th time segment follows
    \begin{equation}\label{eqB0}
        \Pr(n_m^t = \ell) = \sum_{j=1}^J p_{m,j} \frac{(\lambda_j T_s )^{\ell}}{\ell !} e^{-\lambda_j T_s},
    \end{equation}
    where $\lambda_j$ is sampled from $\Lambda$ with probability $\Pr(\lambda _j) = p_{m,j}$ 
    in the $m$-th cell. It is worth noting that the mixed Poisson process can be used to fit arbitrary distributions {\cite{https://doi.org/10.48550/arxiv.1901.06708}}.
    \item The staying time of a user in the $m$-cell follows the exponential distribution with parameter $\mu_m$, where $\mu_m$ is the mean service time of a user.
\end{enumerate}
\end{defi}

Given the user arrival and departure processes, we first analyze the number of users in a cell at the beginning of a time segment $\widetilde n_m^t$.

\begin{prop}[The distribution of residual users]\label{prop_residual_users_distribution}
At the beginning of the $t$-th time segment, the probability mass function (PMF) of the number of users in the $m$-th cell $\widetilde n_m^t$ is given by
\begin{equation}\label{eqB1}
    \Pr(\widetilde n_m^t = \ell) = \sum_{j=1}^J p_{m,j} \frac{\left( \widetilde{\lambda}_{m,j} \right)^{\ell}}{\ell !} e^{-\widetilde{\lambda}_{m,j}},
\end{equation}
for $\ell = 0,1,2,\cdots$, where $\widetilde{\lambda}_{m,j} = \frac{\lambda_j}{\mu_m}(1-e^{-\mu_m T_s})$. The average $\widetilde n_m^t$ is $\mathbb{E} [\widetilde n_m^t] = \sum_{j=1}^J p_{m,j} \widetilde{\lambda}_{m,j} $.
\end{prop}

\begin{NewProof}
See Appendix~\ref{sec:AppA}.
\end{NewProof}

The dynamic power consumption of a cell is proportional to the number of users to be served $N_m^t=\widetilde n_m^t+n_m^t$. At the beginning of the $t$-th time segment, the number of existing users in the cells $\{\widetilde n_m^t: m = 1,2, \cdots ,M\} $ is known to the central controller, while the number of users to arrive in the $t$-th time segment is a random variable, the exact number of which is unknown. Thus, the central controller can only make decisions based on the expected number of users to arrive in the $t$-th time segment.

\begin{prop}[Anticipated power consumption of a cell]\label{prop_power_cell}
Denote by ${\mathcal{P}}_d$ and ${\mathcal{P}}_e$ the average power consumption of the gNB and ng-eNB to serve a single user, respectively. At the beginning of the $t$-th time segment, the anticipated power consumption of the $m$-th cell is given by
\begin{eqnarray}\label{eqB2}
    \hspace{-0.7cm}&& {\mathcal{P}} \left( s_m^t=(a_m^{t-1},\widetilde n_m^t),a_m^t \right)=  \\
    \hspace{-0.7cm}&& \begin{cases}
    {\mathcal{P}}_\text{static} + \left( \widetilde n_m^t + \overline{\lambda}_m T_s \right) {\mathcal{P}}_d,& \text{if}~a_m^t=a_m^{t-1}=1;\\
    {\mathcal{P}}_\text{static} + {\mathcal{P}}_{switch} + \left( \widetilde n_m^t + \overline{\lambda}_m T_s \right) {\mathcal{P}}_d,& \text{if}~a_m^t=1,a_m^{t-1}=0;\\
    \left( \widetilde n_m^t + \overline{\lambda}_m T_s \right) {\mathcal{P}}_e,& \text{if}~a_m^t=0,
    \end{cases}\nonumber
\end{eqnarray}
where $\overline{\lambda}_m \triangleq \sum_{j=1}^J p_{m,j} \lambda_j$ and $\overline{\lambda}_m T_s$ is the expected number of newly arrived users in the $t$-th time segment in the $m$-th cell.
\end{prop}

\section{The Optimal Policy and the Greedy Policy}\label{sec:III}
\subsection{The optimal switching policy}\label{sec:IIIA}
Given the definitions in Section~\ref{sec:II}, the discrete MDP associated with the on-off switching of gNBs can be described as follows. At the beginning of the $t$-th time segment, the central controller observes a system state $\bm{s}^t=(s_1^t,s_2^t, \cdots ,s_M^t)^\top$ and determines a set of configurations $\bm{a}^t=(a_1^t,a_2^t,\cdots,a_M^t)^\top$ for the gNBs following its policy $\pi$. The action produces two results: an immediate cost $C(\bm{s}^t,\bm{a}^t)$ is incurred, and the system evolves to a new state $\bm{s}^{t+1}$ in the next time segment.

The optimal switching policy $\pi^*$ that minimizes the long-term average cost in \eqref{eqA2} satisfies the Bellman equation:
\begin{equation}\label{eqC1}
    g^* + h_{\pi^*} [\bm{s}] = \mathop {\min }_{\bm{a}} \left\{ C(\bm{s},\bm{a}) + \sum_{\bm{s}^\prime} \Pr ( \bm{s}^\prime\left| \bm{s},\bm{a} \right. ) h_{\pi ^*}[\bm{s}^\prime] \right\}
\end{equation}
where $h_{\pi^*} [\bm{s}]$ is the relative value function of a state $\bm{s}$ under the optimal policy $\pi^*$; $g^*$ is the average cost incurred per time step under the optimal policy $\pi^*$; $\Pr (\bm{s}^\prime\left| {\bm{s},\bm{a}} \right.)$ is the probability that the system moves to a new state $\bm{s}^\prime$ when the action $\bm{a}$ is executed in the state $\bm{s}$. Specifically, we have
\begin{eqnarray}\label{eqC2}
    \Pr (\bm{s}^\prime\left| \bm{s},\bm{a} \right.) \hspace{-0.3cm}&\overset{(a)}{=}&\hspace{-0.3cm} \prod_{m = 1}^M \Pr (s^\prime_m \left| s_m,a_m \right.) \\
    &\overset{(b)}{=}&\hspace{-0.3cm} \prod_{m = 1}^M \Pr({\widetilde n^\prime}_m \left| \widetilde{n}_m,a_m \right.) \nonumber\\
    &\overset{(c)}{=}&\hspace{-0.3cm} \prod_{m = 1}^M \left( {\sum_{j = 1}^J { p_{m,j} \frac{{\left( \widetilde{\lambda}_{m,j} \right)}^{{\widetilde n^\prime}_m}}{{\widetilde n^\prime}_m !} e^{ - {\widetilde \lambda }_{m,j}} } } \right) \nonumber,
\end{eqnarray}
where (a) follows from the independent assumption of the cells; (b) follows because the actions are determined once a policy is given -- the only random variable in the state $s^\prime_m$ is ${\widetilde n^\prime}_m$; (c) follows from Proposition~\ref{prop_residual_users_distribution}.

The solution $(g^*,h_{\pi^*})$ to \eqref{eqC1} can be solved by the relative value iteration algorithm (RVIA) since the MDP is unichain {\cite{puterman2014markov}}. Once \eqref{eqC1} is solved, the optimal policy $\pi^*$ can be extracted by acting greedy, i.e., choosing the action that gives the minimal cost:
\begin{equation}\label{eqC3}
    \pi^* (\bm{s}) = \arg \min_{\bm{a}} \left\{ C(\bm{s},\bm{a}) + \sum_{\bm{s}^\prime} \Pr (\bm{s}^\prime \left| \bm{s},\bm{a} \right.) h_{\pi^*} [\bm{s}^\prime] \right\}.
\end{equation}

This solution of RVIA is optimal, but it exhibits several problems. First, the number of states needs to be finite to solve \eqref{eqC1}. In our problem, the state size is infinite as the number of users in a cell can be any non-negative integers. Therefore, an upper limit $N_{th}$ has to be set such that the number of users in a cell is reset to $N_{th}$ if it is larger than $N_{th}$. In order not to affect the optimality of the RVIA, $N_{th}$ should be large enough to make $\Pr(\widetilde n_m^t > N_{th}) $ negligible for a set of policies in the neighborhood of the optimal switching policy.

Given the upper limit $N_{th}$, the state size is $|\bm{s}| = 
\left[\binom{M}{0} + \binom{M}{1} + \cdots + \binom{M}{K} \right] (N_{th}+1)^M$
and the action size is $|\bm{a}| = \binom{M}{0} + \binom{M}{1} + \cdots + \binom{M}{K}$. The decision space is thus
\begin{equation}\label{eqC4}
    |\bm{s}| \!\times\! |\bm{a}| \!\times\! |\bm{s}| \!=\! \left[ \! \binom{M}{0} \!+\! \binom{M}{1} \!+\! \cdots \!+\! \binom{M}{K}\! \right]^3 \! (N_{th} \!+\! 1)^{2M}.
\end{equation}

This implies a formidable computational complexity of RVIA as the complexity will grow exponentially with the increase of $M$ -- often, the optimal policy is incomputable when $M$ is larger than 4. 

\subsection{The greedy switching policy}\label{sec:IIIB}
Considering the high computational complexity of the optimal policy, a widely-used alternative to solve the discrete MDP is the greedy policy \cite{POMDP}. As the name suggests, the greedy policy minimizes the immediate cost of the current time segment, as opposed to the long-term average cost $\overline{C}$. Denote by $\pi_g$ the greedy policy. At the beginning of the $t$-th time segment, the action $\bm{a}_g^t$ chosen by the greedy policy can be written as
\begin{equation}\label{eqD1}
    \bm{a}_g^t = \pi_g (\bm{s}^t) = \arg \min_{\bm{a}^t} C(\bm{s}^t,\bm{a}^t),~\forall t.
\end{equation}

Compared with the optimal policy, the decision criterion of the greedy policy is relatively simpler as it considers only the immediate cost.

In our problem, the decision space of the greedy policy is $|\bm{s}| \times |\bm{a}| = \left[ \binom{M}{0} + \binom{M}{1} + \cdots + \binom{M}{K} \right]^2 (N_{th} + 1)^M $, which also scales with $M$. However, when $K = M$ (i.e., all the gNBs are allowed to be turned off), the greedy policy exhibits a nice threshold structure, as demonstrated in Theorem~\ref{thm_greedy_structure}, and can be computed efficiently. For simplicity, the anticipated cost of a time segment for the $m$-th cell is defined as follows
\begin{eqnarray}\label{eqD5}
    C_m^{(01)} \hspace{-0.3cm} &\triangleq& \hspace{-0.3cm}  \mathbb{E}_{\widetilde n_m^t}\left\{ f\left[ {\mathcal{P}} \left( s_m^t = (a_m^{t-1} = 0,\widetilde n_m^t), a_m^t = 1 \right) \right] \right\} \nonumber\\
    \hspace{-0.3cm} &=& \hspace{-0.3cm} \mathbb{E}_{\widetilde n_m^t}\left\{ f\left[  {\mathcal{P}}_\text{static} + {\mathcal{P}}_\text{switch} + (\widetilde n_m^t + {\overline \lambda }_m T_s) {\mathcal{P}}_d \right] \right\}, \nonumber\\
    C_m^{(11)} \hspace{-0.3cm} &\triangleq& \hspace{-0.3cm} \mathbb{E}_{\widetilde n_m^t}\left\{ f\left[ {\mathcal{P}} \left( s_m^t = (a_m^{t-1} = 1,\widetilde n_m^t), a_m^t = 1 \right) \right] \right\} \nonumber\\
    \hspace{-0.3cm} &=& \hspace{-0.3cm} \mathbb{E}_{\widetilde n_m^t}\left\{ f\left[  {\mathcal{P}}_\text{static} + (\widetilde n_m^t + {\overline \lambda }_m T_s) {\mathcal{P}}_d \right] \right\}, \nonumber\\
    C_m^{(0)} \hspace{-0.3cm} &\triangleq& \hspace{-0.3cm} \mathbb{E}_{\widetilde n_m^t}\left\{ f\left[ {\mathcal{P}} \left( s_m^t =(a_m^{t-1},\widetilde n_m^t),a_m^t = 0 \right) \right] \right\} \nonumber\\
    \hspace{-0.3cm} &=& \hspace{-0.3cm} \mathbb{E}_{\widetilde n_m^t}\left\{ f\left[ (\widetilde n_m^t + {\overline \lambda }_m T_s) {\mathcal{P}}_e \right] \right\},
\end{eqnarray}
where $C_m^{(01)}$ is the anticipated cost of a time segment when $a_m^{t-1}=0$ and $a_m^t=1$; $C_m^{(11)}$ is the anticipated cost of a time segment when $a_m^{t-1}=a_m^t=1$; and $C_m^{(0)}$ is the anticipated cost of a time segment when $a_m^t=0$.

\begin{thm}[Structure of the greedy policy]\label{thm_greedy_structure}
The greedy policy is a dual-threshold policy when $K = M$. Specifically, for the $m$-th cell, $m=1,2,\cdots,M$, there exists two thresholds
\begin{equation}\label{eqD2}
    \gamma _m^L \triangleq \frac{{\mathcal{P}}_\text{static}}{{\mathcal{P}}_e-{\mathcal{P}}_d}-\overline{\lambda}_m T_s,
    \gamma _m^U \triangleq \frac{{\mathcal{P}}_\text{static}+{\mathcal{P}}_\text{switch}}{{\mathcal{P}}_e-{\mathcal{P}}_d}-\overline{\lambda}_m T_s,
\end{equation}
such that the action $a_m^t$ under the greedy policy is given by
\begin{eqnarray}\label{eqD3}
    a_m^t =
    \begin{cases}
        1,& \text{if}~a_m^{t-1}=0~\text{and}~\widetilde{n}_m^t>\gamma_m^U;\\
        0,& \text{if}~a_m^{t-1}=0~\text{and}~\widetilde{n}_m^t\le\gamma_m^U;\\
        1,& \text{if}~a_m^{t-1}=1~\text{and}~\widetilde{n}_m^t>\gamma_m^L;\\
        0,& \text{if}~a_m^{t-1}=1~\text{and}~\widetilde{n}_m^t\le\gamma_m^L.\\
    \end{cases}
\end{eqnarray}

The long-term average cost of the greedy policy is
\begin{eqnarray}\label{eqD4}
    \overline{C}_g \hspace{-0.3cm}&=&\hspace{-0.3cm} \sum_{m=1}^M \left( \frac{p_m^L}{p_m^L+p_m^U} p_m^U C_m^{(01)} \right. \\
    &&\hspace{0.3cm} + \left. \frac{p_m^U}{p_m^L+p_m^U} (1-p_m^L) C_m^{(11)} + \frac{p_m^L}{p_m^L+p_m^U} C_m^{(0)} \right), \nonumber
\end{eqnarray}
where $p_m^L \triangleq \Pr (\widetilde{n}_m^t<\gamma_m^L)$; $p_m^U \triangleq \Pr (\widetilde{n}_m^t>\gamma_m^U)$.
\end{thm}

\begin{NewProof}
When $K=M$, the actions between each cell can be made independently. Therefore, we can focus on the greedy policy in one cell -- the immediate cost of the cell cluster is minimized as long as that of each cell is minimized.

Consider the $m$-th cell. Under the greedy policy, the $m$-th gNB will be turned on in the $t$-th time segment if $c(s_m^t,a_m^t=1)<c(s_m^t,a_m^t=0)$, where $c(s_m^t,a_m^t) = f\left[ {\mathcal{P}} (s_m^t,a_m^t) \right]$ as in Definition~\ref{defi_immediate_cost}. The tie breaks arbitrarily.

Thus, when $a_m^{t-1}=0$, we have
\begin{equation*}
    f\left[ (\widetilde n_m^t \!+\! \overline \lambda _m T_s) {\mathcal{P}}_e \right] \!>\! 
    f\left[ {\mathcal{P}}_\text{static} \!+\! {\mathcal{P}}_\text{switch} \!+\! (\widetilde n_m^t \!+\! \overline \lambda _m T_s) {\mathcal{P}}_d \right],
\end{equation*}
i.e.,
\begin{equation*}
    \widetilde n_m^t > \frac{{\mathcal{P}}_\text{static} + {\mathcal{P}}_\text{switch}}{{\mathcal{P}}_e - {\mathcal{P}}_d} - \overline \lambda _m T_s \triangleq \gamma_m^U.
\end{equation*}

When $a_m^{t-1}=1$, we have
\begin{equation*}
    f\left[ (\widetilde n_m^t + \overline \lambda _m T_s) {\mathcal{P}}_e \right] >  f\left[ {\mathcal{P}}_\text{static} + (\widetilde n_m^t + \overline \lambda _m T_s)  {\mathcal{P}}_d  \right],
\end{equation*}
i.e.,
\begin{equation*}
    \widetilde n_m^t > \frac{{\mathcal{P}}_\text{static}}{{\mathcal{P}}_e - {\mathcal{P}}_d} - \overline \lambda _m T_s \triangleq \gamma_m^L .
\end{equation*}

In other words, depending on whether the gNB is ON or OFF in the $(t-1)$-th time segment, we have two thresholds $\gamma_m^L$ and $\gamma_m^U$ for $\widetilde n_m^t$: the gNB will be turned on if $\widetilde n_m^t$ is larger than the two thresholds. On the other hand, the gNB will be turned off if $\widetilde n_m^t$ is smaller than the two thresholds when the gNB is ON and OFF in the $(t-1)$-th time segment, respectively. This gives us the dual-threshold structure of the greedy policy.

In the above context, the state transitions of a single cell under the greedy policy can be viewed as a Markov chain, and the state transition matrix is given by
\begin{eqnarray}\label{eqD7}
    \begin{bmatrix}
    {1 - p_m^L} & {p_m^L}\\ {p_m^U} & {1 - p_m^U}
    \end{bmatrix},
\end{eqnarray}
where $p_m^L \triangleq \Pr (\widetilde{n}_m^t<\gamma_m^L)$; $p_m^U \triangleq \Pr (\widetilde{n}_m^t>\gamma_m^U)$. The stationary distribution of the Markov chain is $\left( \frac{p_m^U}{p_m^L+p_m^U} , \frac{p_m^L}{p_m^L+p_m^U} \right)$, where the two probabilities correspond to ON and OFF, respectively.

As a result, for a single cell, the long-term average cost of the greedy policy can be computed by
\begin{eqnarray*}
    \overline C_{g,m} \hspace{-0.3cm}&=&\hspace{-0.3cm} \Pr(a_m^{t-1}=0,a_m^t=1) C_m^{(01)} + \\
    && \hspace{-0.3cm} \Pr (a_m^{t-1}=1,a_m^t=1) C_m^{(11)} + \Pr (a_m^t=0) C_m^{(0)} \\
    &=& \hspace{-0.3cm} \frac{p_m^L}{p_m^L+p_m^U} p_m^U C_m^{(01)} + \frac{p_m^U}{p_m^L+p_m^U} (1-p_m^L) C_m^{(11)} + \\
    && \hspace{-0.3cm} \frac{p_m^L}{p_m^L+p_m^U} C_m^{(0)},
\end{eqnarray*}
where $C_m^{(01)}$, $C_m^{(11)}$, and $C_m^{(0)}$ are as defined in \eqref{eqD5}. The long-term average cost of the cell cluster is $\overline C_g = \sum_{m=1}^M \overline C_{g,m}$ as \eqref{eqD4}.
\end{NewProof}


Next, we analyze the connections between the greedy policy and the optimal policy.

\begin{prop}[Connections between the greedy policy and the optimal policy]\label{prop_greedy_optimal}
Let $K=M$. For any cell in the cluster,
\begin{enumerate}
    \item If the optimal policy instructs the gNB to turn off, the action of the greedy policy is also off.
    \item If the greedy policy instructs the gNB to turn on, the optimal action is also on.
\end{enumerate}
\end{prop}

\begin{NewProof}
We prove Proposition~\ref{prop_greedy_optimal} by contradiction.
With the optimal policy $\pi^*$, the long-term cost of the $m$-th cell is given by
\begin{equation*}
    \sum_{t=1}^{\infty} {c_{\pi^*} (s_m^t,a_m^t)} .
\end{equation*}

At a time segment $t_0$, suppose the optimal action is OFF and the greedy action is ON.
First, we have $c(s_m^{t_0}, a_m^{t_0}=1)<c(s_m^{t_0},a_m^{t_0}=0)$ since the greedy action is ON.
We can then construct a new policy as follows: i) in the $t_0$-th time segment, the new policy instructs the gNB to turn on, and ii) in any other time segments, it has the same action as the optimal policy.
As can be seen, the cost incurred by the new policy differs from that of the optimal policy only in the $t_0$-th and $(t_0+1)$-th time segments.
In particular, the new policy incurs lower costs than the optimal policy since
\begin{eqnarray*}
    &&\hspace{-0.3cm} c_{\pi^*} (s_m^{t_0}, a_m^{t_0}=0) + c_{\pi^*} (s_m^{t_0+1}, a_m^{t_0+1}) \\
    = \hspace{-0.3cm}&&\hspace{-0.3cm} c_{\pi^*} (s_m^{t_0}, a_m^{t_0}=0) +\\
    &&\hspace{-0.3cm} \begin{cases}
        f [ (\widetilde n_m^{t_0+1} + \overline \lambda _m T_s) {\mathcal{P}}_e ], &\hspace{-0.2cm} \text{if}~a_m^{t_0+1}=0;\\
        f [{\mathcal{P}}_\text{static} \!+\! {\mathcal{P}}_\text{switch} \!+\! (\widetilde n_m^{t_0+1} \!+\! \overline \lambda _m T_s) {\mathcal{P}}_d],&\hspace{-0.2cm} \text{if}~a_m^{t_0+1}=1;
    \end{cases} \\
    > \hspace{-0.3cm}&&\hspace{-0.3cm} c_{\pi^*} (s_m^{t_0}, a_m^{t_0}=1) +\\
    &&\hspace{-0.3cm} \begin{cases}
        f [ (\widetilde n_m^{t_0+1} + \overline \lambda _m T_s) {\mathcal{P}}_e ], & \text{if}~a_m^{t_0+1}=0;\\
        f [{\mathcal{P}}_\text{static} \!+\! (\widetilde n_m^{t_0+1} \!+\! \overline \lambda _m T_s) {\mathcal{P}}_d],& \text{if}~a_m^{t_0+1}=1,
    \end{cases}
\end{eqnarray*}
yielding a contradiction. As a result, for any time segments, if the optimal action is OFF, the greedy action must also be OFF.
    
Likewise, it can be proven that when the greedy action is ON, the optimal action must also be ON; otherwise, a contradiction occurs.
\end{NewProof}

When $K=M$, Proposition~\ref{prop_greedy_optimal} indicates that the number of ``ON'' time segments under the greedy policy is no greater than that under the optimal policy.


\section{The Index Policy}\label{sec:IV}
The optimal policy and the greedy policy are computationally expensive because the configurations of the $M$ cells are coupled together, yielding decision spaces growing exponentially with $M$.
One exception is the greedy policy with $K = M$, in which case the central controller can configure each cell independently and the greedy policy exhibits a dual-threshold structure.
This implies that decoupling the cell configurations is a key to devising computationally efficient policies.

Decoupling the MDP dates back to the Whittle index approach to solve the RMAB problem \cite{whittle_1988}. The general idea is to decouple the $M$-dimensional MDP to $M$ one-dimensional MDPs, each decoupled MDP can then be solved efficiently thanks to the largely reduced state and action spaces. Inspired by Whittle's approach \cite{whittle_1988}, this paper puts forth an index policy for the on-off switching problem of gNBs.

\subsection{The decoupled problem}\label{sec:IVA}
To start with, we formulate the decoupled problem of (P1), i.e., the on-off switching of a single cell as opposed to the cell cluster. Specifically, we omit the constraint $\sum_{m=1}^M {a_m^t} \le M-K$ in \eqref{eqA2} so that the on-off configurations of gNBs are independent. For the decoupled problem, the subscript $m$ is omitted in this subsection to ease exposition.

The on-off switching of a single cell is a controlled Markov process defined as follows.

\begin{defi}[The decoupled problem]\label{defi_decoupled_problem}
Consider a single cell. The state of the cell at the beginning of the $t$-th time segment is $s^t=(a^{t-1}, \widetilde{n}^t)$, where $a^{t-1}$ is the ON/OFF state of the cell in the $(t-1)$-th time segment and $\widetilde{n}^t \in \mathbb{N}$ stands for the number of users in the cell at the beginning of the $t$-th time segment. The immediate cost incurred by a state-action pair $\left( {s^t=(a^{t-1},\widetilde{n}^t),a^t} \right)$ is 
\begin{eqnarray}\label{eqE1}
    \hspace{-0.7cm} && c(s^t,a^t) = \\
    \hspace{-0.7cm} && \begin{cases}
        f\left[ {\mathcal{P}}_\text{static}+(\widetilde{n}^t+\overline{\lambda} T_s) {\mathcal{P}}_d \right], &\hspace{-0.2cm} \text{if}~a^t=a^{t-1}=1; \\
        f\left[ {{\mathcal{P}}_\text{static}+{\mathcal{P}}_\text{switch}+( \widetilde{n}^t+\overline{\lambda} T_s ) {\mathcal{P}}_d } \right], &\hspace{-0.2cm} \text{if}~a^t=1,a^{t-1}=0;\\
        f\left[ {( \widetilde{n}^t+\overline{\lambda} T_s ) {\mathcal{P}}_e} \right] + \epsilon, &\hspace{-0.2cm} \text{if}~a^t=0,
    \end{cases} \nonumber
\end{eqnarray}
where $\epsilon$ is a cost of being OFF (which will be explained later). The optimal policy $\widetilde \pi ^*$ for the decoupled problem is defined as
\begin{equation}\label{eqE2}
    (\text{P2}): \widetilde \pi ^* = \arg {\min _{\widetilde \pi} {\lim _{T \to \infty} {\mathbb{E} \left[ \frac{1}{T} \sum_{t=0}^{T-1} {c(s^t,a^t)} \right]}} }.
\end{equation}
\end{defi}

Compared with the original $M$-dimensional MDP, the decoupled problem introduces a cost $\epsilon$ in \eqref{eqE1} to penalize the OFF action. In particular, we aim to find the critical cost $\epsilon^*$ for each state such that the expected costs incurred by turning on and off at the state are the same. In doing so, the index $\epsilon^*$ acts as a measurement of how the controller is willing to pay to turn off the gNB. Said in another way, $\epsilon^*>0$ means that the average cost for the action $a=0$ (without the cost $\epsilon^*$) is smaller than that when $a=1$: the controller prefers to turn off this station, the larger $\epsilon^*$ is, the more likely it is to be turned off. On the other hand, $\epsilon^*<0$ means that the average cost for the action $a=1$ is smaller, so the gNB is likely to be turned on.

Given the above analysis, in the original problem with $M$ cells, we can compute the corresponding indexes $\epsilon^*$ for individual cells in each time segment according to their states.
If more than $K$ gNBs have positive $\epsilon^*$, then we turn off the gNBs with $K$ largest indexes.
If less than $K$ gNBs have positive $\epsilon^*$, then we only turn off those with positive indexes.

Next, we study how to solve the decoupled problem. As explained in Section~\ref{sec:IIIA}, the optimal policy of an MDP follows the Bellman equation. For the decoupled problem, the optimal policy $\widetilde \pi ^*$ follows
\begin{equation}\label{eqE3}
    g^* + h_{\widetilde \pi ^*} [s] = \min \left\{ {Q(s,a=1),Q(s,a=0)} \right\},
\end{equation}
where 
\begin{eqnarray*}\label{eqE4}
    && Q(s,a=1) = c(s,a=1)+\Sigma_1 ,\\
    && Q(s,a=0) = c(s,a=0)+\Sigma_0+\epsilon , \\
    && \Sigma_1 \triangleq \sum_{k=0}^{\infty} {\Pr(\widetilde{n} = k) \cdot h_{\widetilde \pi ^*}[s=(1,k)]} ,\\
    && \Sigma_0 \triangleq \sum_{k=0}^{\infty} {\Pr(\widetilde{n} = k) \cdot h_{\widetilde \pi ^*}[s=(0,k)]},
\end{eqnarray*}
and $\Pr(\widetilde{n} = k)$ represents the distribution of the residual users, as per \eqref{eqB1}; $h_{\widetilde \pi ^*} [s]$ is the relative value function of a state $s$ under the optimal policy $\widetilde \pi ^*$; $c(s,a)$ is the immediate cost given in \eqref{eqE1}; $g^*$ is gain of the decoupled MDP.

To be more specific, we write one iteration of \eqref{eqE3} in the equilibrium as follows:
\begin{eqnarray}\label{eqE5}
    &&\hspace{-0.8cm} Q[s=(1,\widetilde{n}), a=1] = f\left[ {\mathcal{P}}_\text{static}+ (\widetilde{n} + \overline \lambda T_s) {\mathcal{P}}_d \right]+\Sigma_1 \nonumber ,\\
    &&\hspace{-0.8cm} Q[s=(1,\widetilde{n}), a=0] = f\left[ (\widetilde{n} + \overline \lambda T_s) {\mathcal{P}}_e \right]+\Sigma_0+\epsilon \nonumber , \\
    &&\hspace{-0.8cm} h_{\widetilde \pi ^*} [s=(1,\widetilde{n})] = \nonumber \\
    &&\hspace{-0.5cm} \min \left\{ Q[s=(1,\widetilde{n}), a=1],~Q[s=(1,\widetilde{n}), a=0] \right\} \!-\! g^* \nonumber ,\\
    &&\hspace{-0.8cm} Q[s=(0,\widetilde{n}), a=1] \!=\! f\left[ {\mathcal{P}}_\text{static} \!+\! {\mathcal{P}}_\text{switch} \!+\! (\widetilde{n} \!+\! \overline \lambda T_s) {\mathcal{P}}_d \right] \!+\! \Sigma_1 \nonumber ,\\
    &&\hspace{-0.8cm} Q[ s=(0,\widetilde{n}), a=0] = f\left[ (\widetilde{n} + \overline \lambda T_s) {\mathcal{P}}_e \right]+\Sigma_0+\epsilon \nonumber , \\
    &&\hspace{-0.8cm} h_{\widetilde \pi ^*} \left[s=(0,\widetilde{n})\right] = \nonumber \\
    &&\hspace{-0.5cm} \min \left\{ Q[s=(0,\widetilde{n}), a=1],~Q[ s=(0,\widetilde{n}), a=0] \right\} \!-\! g^*.
\end{eqnarray}

Without loss of generality, we choose state $s=(1,0)$ as the reference state and set $h_{\widetilde{\pi}^*}\left[s=(1,0)\right]=0$. Eq. \eqref{eqE5} defines the relative value function of each state under the optimal policy for a given cost $\epsilon$.

Intuitively, the gNB of a cell has to be turned on when the number of users in the cell is large, and turned off otherwise.
Inspired by the dual-threshold structure of the greedy policy when $K=M$, a natural question is that, does the decoupled MDP exhibits a threshold structure?
In the following, we answer this question affirmatively by proving that the dual-threshold structure of the optimal policy $\widetilde{\pi}^*$ to the problem (P2).

\begin{prop}[Structure of the optimal policy $\widetilde \pi ^*$]\label{prop_decoupled_solution}
The optimal policy $\widetilde \pi ^*$ for the decoupled problem (P2) is a dual-threshold policy. For a given cost $\epsilon$, there exists two thresholds $\Gamma^L$, $\Gamma^U$, and $\Gamma^L<\Gamma^U$, such that the optimal action $a^t$ is given by
\begin{eqnarray}\label{eqE6}
    a^t =
    \begin{cases}
        1,& \text{if}~a^{t-1}=0~\text{and}~\widetilde{n}^t>\Gamma^U;\\
        0,& \text{if}~a^{t-1}=0~\text{and}~\widetilde{n}^t\le\Gamma^U;\\
        1,& \text{if}~a^{t-1}=1~\text{and}~\widetilde{n}^t>\Gamma^L;\\
        0,& \text{if}~a^{t-1}=1~\text{and}~\widetilde{n}^t\le\Gamma^L.
    \end{cases}
\end{eqnarray}
\end{prop}

\begin{NewProof}
Given a fixed $\epsilon$, $h_{\widetilde{\pi}^*}\left[s=(1,\widetilde{n})\right]$, $h_{\widetilde{\pi}^*}\left[s=(0,\widetilde{n})\right]$, $\Sigma_1$, and $\Sigma_0$ in \eqref{eqE5} are constant when RVIA converges. Under the optimal policy, the gNB will be turned on deterministically in state $s=(1,\widetilde{n})$ if $Q\left[s=(1,\widetilde{n}), a=1 \right] < Q\left[s=(1,\widetilde{n}), a=0 \right]$. As per \eqref{eqE5}, we have 
\begin{equation*}\label{eqE7}
    f\left[ {\mathcal{P}}_\text{static}+ (\widetilde{n}+\overline \lambda T_s) {\mathcal{P}}_d \right]+\Sigma_1 < f\left[ (\widetilde{n}+\overline \lambda T_s) {\mathcal{P}}_e \right]+\Sigma_0+\epsilon .
\end{equation*}
After some manipulation, we have
\begin{equation*}\label{eqE8}
    g^L(\widetilde{n}) > - \epsilon + \Sigma_1 - \Sigma_0,
\end{equation*}
where
$$g^L(\widetilde{n})\! \triangleq \! f\left[ (\widetilde{n}\!+\!\overline \lambda T_s) {\mathcal{P}}_e \right] \!-\! f\left[ {\mathcal{P}}_\text{static}\!+\! (\widetilde{n} \!+\! \overline \lambda T_s) {\mathcal{P}}_d \right]$$ is a monotonically increasing function with $\widetilde{n}$. As a result, there exists a threshold $\Gamma^L$, for a given state $s=(1,\widetilde{n})$, if $\widetilde{n} > \Gamma^L$, the optimal action is to turn on the gNB, otherwise, it should be turned off, and the threshold $\Gamma^L$ is the solution of the following equation:
\begin{equation}\label{eqE9}
    g^L(\Gamma^L) = - \epsilon + \Sigma_1 - \Sigma_0.
\end{equation}

Likewise, for a given state $s=(0,\widetilde{n})$, the gNB will be turned on if $Q\left[s=( 0,\widetilde{n}), a=1 \right] < Q\left[ s=(0,\widetilde{n}), a=0 \right]$. As per \eqref{eqE5}, it can be written as
\begin{equation*}\label{eqE10}
    g^U(\widetilde{n}) > - \epsilon + \Sigma_1 - \Sigma_0,
\end{equation*}
where
\begin{equation*}\label{eqE11}
    g^U(\widetilde{n})\! \triangleq \! f\left[ ( \widetilde{n}\!+\!\overline \lambda T_s ) {\mathcal{P}}_e \right] \!-\! f\left[ {\mathcal{P}}_\text{static}\!+\! {\mathcal{P}}_\text{switch}\!+\!(\widetilde{n}\!+\!\overline \lambda T_s) {\mathcal{P}}_d \right]
\end{equation*}
is also a monotonically increasing function with $\widetilde{n}$, thus the threshold $\Gamma^U$ exists. For the state $s=(0,\widetilde{n})$, if $\widetilde{n} > \Gamma^U$, the optimal action is to turn on the gNB, otherwise, it should be turned off, and the threshold $\Gamma^U$ is the solution of the following equation:
\begin{equation}\label{eqE12}
    g^U(\Gamma^U) = - \epsilon + \Sigma_1 - \Sigma_0.
\end{equation}

In particular, for a given $\epsilon$, we have
\begin{equation}\label{eqE13}
    g^L(\Gamma^L) = g^U(\Gamma^U)
\end{equation}
from \eqref{eqE9} and \eqref{eqE12}.
According to the definitions of $g^L(\cdot)$ and $g^U(\cdot)$, $g^L(\widetilde{n}) > g^U(\widetilde{n})$. Therefore,
$$g^U(\Gamma^U)=g^L(\Gamma^L)>g^U(\Gamma^L),$$
and hence, $\Gamma^L< \Gamma^U$.
\end{NewProof}

\begin{figure*}
    \centering
    \begin{subfigure}[t]{0.34\linewidth}
        \centering
        \includegraphics[width=\linewidth]{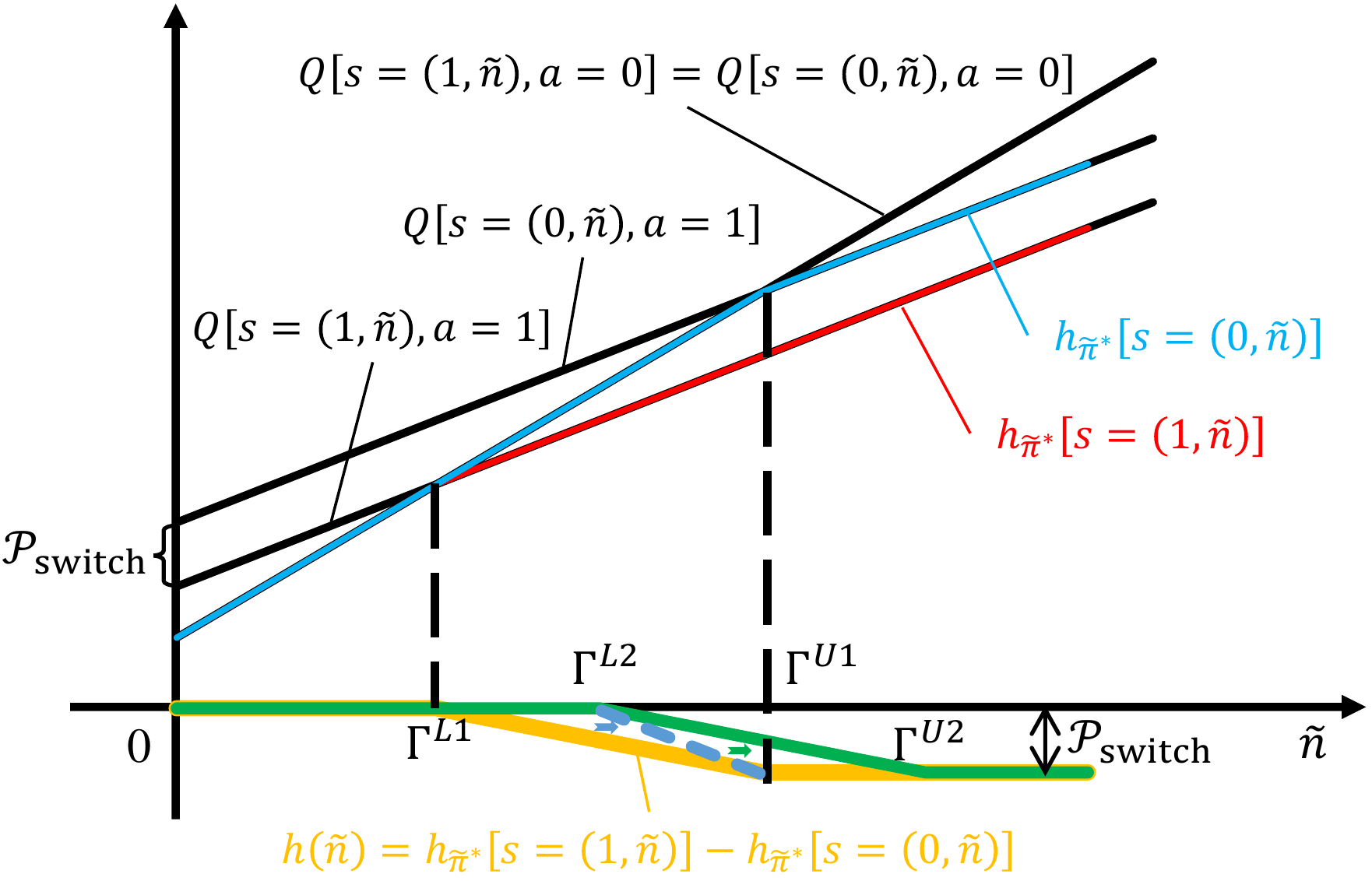}
        \caption{$f(x)=x$.}
    \end{subfigure}
    \qquad
    \begin{subfigure}[t]{0.34\linewidth}
        \centering
        \includegraphics[width=\linewidth]{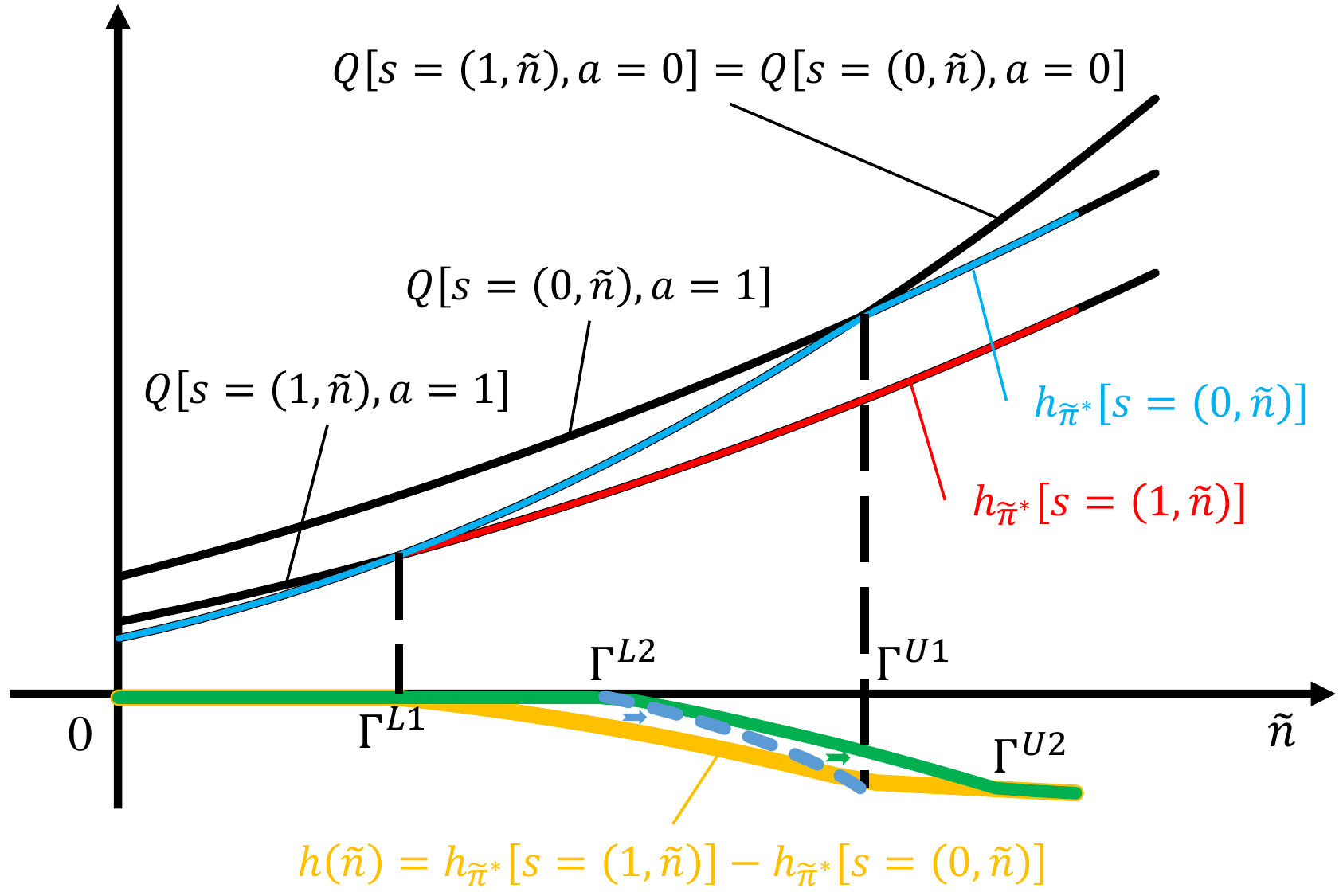}
        \caption{$f(x)=x^2$.}
    \end{subfigure}\\
    \caption{Two examples to illustrate the monotonicity of $h(\widetilde{n})$, where we set $f=x$ and $f=x^2$ in (a) and (b), respectively.}
    \label{fig:example_h}
\end{figure*}

\subsection{The index switching policy}\label{sec:IVB}
Given Proposition \ref{prop_decoupled_solution}, we can define a ``passive set'' that contains all states in which the optimal action is OFF. Specifically, we define
\begin{eqnarray}\label{eqF1}
    \mathcal{D}( \epsilon )\hspace{-0.6cm} && \triangleq \left\{ s=(0,\widetilde{n}) : 0 \le \widetilde{n} \le \Gamma^U (\epsilon), \right. \\
    \hspace{-0.6cm} && \text{and} \left. s=(1,\widetilde{n}) : 0 \le \widetilde{n} \le \Gamma^L (\epsilon) \right\}.\nonumber
\end{eqnarray}
Note that $\Gamma^U$ and $\Gamma^L$ are functions of $\epsilon$, because they are determined under a given $\epsilon$.

According to \cite{whittle_1988} the decoupled problem is indexable if the passive set $\mathcal{D}(\epsilon)$ defined in \eqref{eqF1} is monotonically non-increasing as $\epsilon$ increases. That is, for any $\epsilon_1<\epsilon_2, (\epsilon_1,\epsilon_2 \in \mathbb{R})$, the passive set $\mathcal{D}(\epsilon_1) \subseteq \mathcal{D}(\epsilon_2)$. The original problem is indexable if all its decoupled problems are indexable. In the following, we shall prove that the decoupled problem (P2) is indexable, and propose an algorithm to compute the index $\epsilon$ for each state.

Given \eqref{eqF1}, the passive set $\mathcal{D}(\epsilon)$ is monotonically non-increasing if and only if both $\Gamma^U(\epsilon)$ and $\Gamma^L(\epsilon)$  decrease monotonically as $\epsilon$ increases.
To prove this, we first establish the monotonicity of $H(\Gamma^L$, $\Gamma^U)\triangleq \Sigma_1 - \Sigma_0$ in the following.

\begin{lem}\label{lem_H_monotone_nondecreasing}
$H(\Gamma^L,\Gamma^U)= \Sigma_1 - \Sigma_0$ is a monotonically non-decreasing function with the increase of $\Gamma^L$ and $\Gamma^U$.
\end{lem}
\begin{NewProof}
To start with, let us define $h(\widetilde{n}) \triangleq h_{\widetilde{\pi}^*}\left[ s=(1,\widetilde{n} )\right]-h_{\widetilde{\pi}^*}\left[s=( 0,\widetilde{n})\right]$. It can further be written as
\begin{eqnarray}\label{eqF2}
    h(\widetilde{n})=
    \begin{cases}
        0,& \hspace{-0.3cm} \widetilde{n} \le \Gamma^L (\epsilon); \\
        \Delta_1 (\widetilde{n}),& \hspace{-0.3cm} \widetilde{n} > \Gamma^U (\epsilon); \\
        \Delta_2 (\widetilde{n}) + \Sigma_1 - \Sigma_0 - \epsilon, & \hspace{-0.3cm} \Gamma^L (\epsilon) < \widetilde{n} \le \Gamma^U (\epsilon) ,
    \end{cases}
\end{eqnarray}
where
\begin{eqnarray*}\label{eqF3}
    \Delta_1 (\widetilde{n}) \triangleq \hspace{-0.3cm}&&\hspace{-0.3cm} f\left[ {{\mathcal{P}}_\text{static}+(\widetilde{n}+\overline{\lambda} T_s) {\mathcal{P}}_d } \right] -\\
    &&\hspace{-0.3cm} f\left[ {{\mathcal{P}}_\text{static}+{\mathcal{P}}_\text{switch}+( \widetilde{n}+\overline{\lambda} T_s ) {\mathcal{P}}_d } \right],\\
    \Delta_2 (\widetilde{n}) \triangleq \hspace{-0.3cm}&&\hspace{-0.3cm} f\left[ {{\mathcal{P}}_\text{static}+( \widetilde{n}+\overline{\lambda} T_s ) {\mathcal{P}}_d } \right] - f\left[ {( \widetilde{n}+\overline{\lambda} T_s ) {\mathcal{P}}_e } \right].
\end{eqnarray*}
Note that both $\Delta_1(\widetilde{n})$ and $\Delta_2(\widetilde{n})$ are monotonically non-increasing functions, and $\Delta_1(\widetilde{n}) \le 0$. Thus, $h(\widetilde{n})$ is monotonically non-increasing with $\widetilde{n}$, and $h(\widetilde{n}) \le 0$. 

$H(\Gamma^L,\Gamma^U)$ can be written as
\begin{eqnarray}\label{eqF4}
    H \hspace{-0.3cm}&=&\hspace{-0.3cm} \Sigma_1 - \Sigma_0 \nonumber \\
    &=&\hspace{-0.3cm} \sum_{k=0}^{\infty} {\Pr(\widetilde{n} = k) \cdot \left\{ h_{\widetilde{\pi}^*}\left[ s=(1,k) \right] - h_{\widetilde{\pi}^*}\left[ s=(0,k) \right] \right\} } \nonumber \\
    &=&\hspace{-0.3cm} \sum_{k=0}^{\infty} {\Pr(\widetilde{n} = k) \cdot h(k) }.
\end{eqnarray}

Appendix~\ref{sec:AppB} proves that $h(\widetilde{n})$ is monotonically non-decreasing in $\Gamma^L$ and $\Gamma^U$. Thus,
\begin{eqnarray*}\label{eqF9}
    &&\hspace{-0.5cm} H(\Gamma^{L1},\Gamma^U)-H(\Gamma^{L2},\Gamma^U)\\
    = &&\hspace{-0.5cm} \sum_{k=0}^{\infty} {\Pr(\widetilde{n} = k) \cdot \left[ h(k,\Gamma^{L1},\Gamma^U)-h(k,\Gamma^{L2},\Gamma^U) \right] } \le 0,
\end{eqnarray*}
when $\Gamma^{L1}<\Gamma^{L2}$, i.e., $H(\Gamma^L,\Gamma^U)$ is a monotonically non-increasing function in $\Gamma^L$. And
\begin{eqnarray*}\label{eqF12}
    &&\hspace{-0.5cm} H(\Gamma^L,\Gamma^{U1})-H(\Gamma^L,\Gamma^{U2})\\
    = &&\hspace{-0.5cm} \sum_{k=0}^{\infty} {\Pr(\widetilde{n} = k) \cdot \left[ h(k,\Gamma^L,\Gamma^{U1})-h(k,\Gamma^L,\Gamma^{U2}) \right] } \le 0,
\end{eqnarray*}
when $\Gamma^{U1}<\Gamma^{U2}$, i.e., $H(\Gamma^L,\Gamma^U)$ is a monotonically non-increasing function in $\Gamma^U$.
\end{NewProof}

To illustrate the monotonicity of $H(\Gamma^L$, $\Gamma^U)$ established in Lemma~\ref{lem_H_monotone_nondecreasing}, two examples are given in Fig.~\ref{fig:example_h} with $f=x$ and $f=x^2$, respectively.
For the linear function in Fig.~\ref{fig:example_h}(a),
$Q[s,a=0]$ and $Q[s,a=1]$ are linear in $\widetilde{n}$, as per \eqref{eqE5}, and $Q[s=(1,\widetilde{n}), a=0]-Q[s=(1,\widetilde{n}), a=1]={\mathcal{P}}_\text{switch}$.
Therefore, both the state value $h_{\widetilde \pi ^*}[s]$ and
$h(\widetilde{n})$ defined in \eqref{eqF2} are piecewise linear.
When we increase the two thresholds from $(\Gamma^{L1},\Gamma^{U1})$ to $(\Gamma^{L2},\Gamma^{U2})$, $h(\widetilde{n})$ is monotonically non-decreasing, hence $H(\Gamma^L,\Gamma^U)$ is monotonically non-decreasing.
The same results can be observed from Fig.~\ref{fig:example_h}(b).

Next, we derive upper and lower bounds for $H(\Gamma^L$, $\Gamma^U)$ based on its monotonicity.

\begin{lem}\label{lem_H_bounds}
$H(\Gamma^L,\Gamma^U)$ is bounded by
\begin{equation}\label{eqF13}
    E \le H(\Gamma^L,\Gamma^U) \le 0,
\end{equation}
where $E \triangleq \sum_{k=0}^{\infty} {\Pr (\widetilde{n} = k)  \Delta_1 (k)}$ is a constant and $\Delta_1 (k)$ is defined in \eqref{eqF2}.
\end{lem}

\begin{NewProof}
From the definition of $h(\widetilde{n})$ in \eqref{eqF2}, we have $h(\widetilde{n}) \le 0$, $\forall \widetilde{n} \ge 0$. Thus, $H(\Gamma^L,\Gamma^U) \leq 0 $, as per \eqref{eqF4}. Since $H(\Gamma^L,\Gamma^U)$ is monotonically non-decreasing with the increase of $\Gamma^L$ and $\Gamma^U$, we have
\begin{eqnarray}\label{eqF14}
H(\Gamma^L,\Gamma^U) \hspace{-0.3cm}&\ge&\hspace{-0.3cm} {H(\Gamma^L,0)} = {\sum_{k=0}^{\infty} {\Pr( \widetilde{n} = k) \cdot h(k)}} \nonumber \\
\hspace{-0.3cm}&\overset{(a)}{=}&\hspace{-0.3cm}  {\sum_{k=0}^{\infty} {\Pr(\widetilde{n} = k) \cdot \Delta_1 (k)}} \triangleq E,
\end{eqnarray}
where $E$ is a constant; $(a)$ holds because the gNB is ON in all states when $\Gamma^U \le 0$, and hence,  $h(\widetilde{n})=\Delta_1(\widetilde{n}), \forall \widetilde{n}$, according to \eqref{eqF2}.
\end{NewProof}

Given the above analysis, we are ready to prove the indexability of the decoupled problem.

\begin{thm} \label{lem_Gamma_monotone_decreasing}
The decoupled problem (P2) is indexable.
\end{thm}

\begin{NewProof}
(sketch) The decoupled problem (P2) is indexable if and only if the passive set $\mathcal{D}( \epsilon )$ in \eqref{eqF1} is monotonically non-increasing in $\epsilon$.
Based on \eqref{eqE13}, consider any two costs $\epsilon_1 \ne \epsilon2$.
If $\Gamma^L(\epsilon_1)>\Gamma^L(\epsilon_2)$, then $g^L\left[ \Gamma^L(\epsilon_1) \right] > g^L\left[ \Gamma^L(\epsilon_2) \right]$ and $g^U\left[ \Gamma^U(\epsilon_1) \right] > g^U\left[ \Gamma^U(\epsilon_2) \right]$, hence $\Gamma^U(\epsilon_1)>\Gamma^U(\epsilon_2)$. In contrast, if $\Gamma^L(\epsilon_1)<\Gamma^L(\epsilon_2)$, we have $\Gamma^U(\epsilon_1)<\Gamma^U(\epsilon_2)$. This means that $\Gamma^L(\epsilon)$ and $\Gamma^U(\epsilon)$ have the same monotonicity in $\epsilon$.
Therefore, we only need to prove that $\Gamma^U(\epsilon)$ is monotonically non-increasing. Detailed proof is given in Appendix~\ref{sec:AppC}.
\end{NewProof}

So far, we have established the indexability of the decoupled problem.
The only issue left is how to compute the index for each state.

For a given state $s^t=(a^{t-1},\widetilde{n}^t)$, the index $\epsilon^*(s^t)$ corresponds to the cost of turning off the gNB, and can be computed by setting
$Q(s,a=1)= Q(s,a=0)$, i.e.,
\begin{equation*}
c(s,a=1)+\Sigma_1 = c(s,a=0)+\Sigma_0+\epsilon^*(s).
\end{equation*}
More specifically, for the class of states $s=(1,\widetilde{n})$, the index $\epsilon^*[s=(1,\widetilde{n})]$ corresponds to the case $\Gamma^L(\epsilon^*) = \widetilde{n}$;
for the class of states $s=(0,\widetilde{n})$, on the other hand, the index $\epsilon^*[s=(0,\widetilde{n})]$ corresponds to the case $\Gamma^U(\epsilon^*) = \widetilde{n}$.
Therefore, the index of a state can be obtained by solving $\Gamma^L(\epsilon^*) = \widetilde{n}$ or $\Gamma^U(\epsilon^*) = \widetilde{n}$.

Deriving the closed-form $\epsilon$ for each state is challenging as the two thresholds $\Gamma^L$ and $\Gamma^U$ do not have explicit expressions. Thus, we resort to a numerical approach to compute $\epsilon$ in the following.

Define
\begin{equation}
\Gamma(\epsilon)\triangleq\begin{cases}
\Gamma^L(\epsilon), & \text{if}~ s=(1,\widetilde{n}), \\
\Gamma^U(\epsilon), & \text{if}~ s=(0,\widetilde{n}),
\end{cases} 
\end{equation}

\begin{equation}
F(\epsilon) \triangleq \frac{1}{2}\left[ \Gamma(\epsilon) - \widetilde{n} \right]^2.
\end{equation}
Then, the index $\epsilon$ of a state $s$ is the optimal $\epsilon$ that minimizes $F(\epsilon)$.
According to Theorem~\ref{lem_Gamma_monotone_decreasing}, $\Gamma^L(\epsilon)$ and $\Gamma^U(\epsilon)$ are monotonically decreasing in $\epsilon$, hence the local minimum of $F(\epsilon)$ is also the global minimum. 
The optimal $\epsilon$ can then be found by gradient descent, where the gradient of $F(\epsilon)$ can be approximated by
\begin{equation}\label{eqF27}
    \frac{dF(\epsilon)}{d\epsilon} = \left(\Gamma(\epsilon) - \widetilde{n} \right) \frac{d\Gamma(\epsilon)}{d\epsilon} \approx \widetilde{n}-\Gamma(\epsilon),
\end{equation}
where the approximation follows because $\Gamma(\epsilon)$ is monotonically decreasing according to Theorem~\ref{lem_Gamma_monotone_decreasing}. Given \eqref{eqF27}, The gradient descent approach to compute the index is summarized in Algorithm~\ref{algo:GD}.


\begin{algorithm}[t]
\caption{Index computation via gradient descent.}
\label{algo:GD}
\begin{algorithmic}
\\ \textbf{Input:} $s^t=(a^{t-1},\widetilde{n}^t)$
\State Pick an initial value $\epsilon_0$.
\State Set an error $\xi>0$ and a step size $\beta>0$.
\State $k \gets 0$
\State $\Gamma(\epsilon_k) \gets \widetilde{n}^t$
\State $F(\epsilon_k)=\xi+1$
\While{$F(\epsilon_k) > \xi$}
\State $\epsilon_{k+1} = \epsilon_k-\beta \cdot \left(\widetilde{n} - \Gamma(\epsilon_k) \right)$
\State Calculate $h_{\widetilde \pi ^*}$ by RVIA for the decoupled problem using $\epsilon_{k+1}$.
\State Calculate $\Gamma(\epsilon_{k+1})$ by \eqref{eqE9} or \eqref{eqE12}.
\State $F(\epsilon_{k+1}) \gets \frac{1}{2}\left[ \Gamma(\epsilon_{k+1}) - \widetilde{n} \right]^2$
\State $k \gets k+1$
\EndWhile
\State $\epsilon^*(s^t) = \epsilon_k$
\\ \textbf{Output:} $\epsilon^*(s^t)$
\end{algorithmic}
\end{algorithm}

The index policy for the original $M$-dimensional problem (P1) can be summarized as follows.

\begin{defi}[The index policy]
At the beginning of the $t$-th time segment, the state of the cell cluster $\bm{s}^t= \{s_1^t,\allowbreak s_2^t,\allowbreak \cdots,\allowbreak s_M^t\}$.
With the index policy $\pi_{\text{ind}}$, an index is first computed for each cell: $\epsilon^*(s_m^t)$, $m=1,2,...,M$. The actions are given by
\begin{equation}\label{eqF28}
    \bm{a}_{\text{ind}}^t = \pi_{\text{ind}} (\bm{s}^t) = (a_{\text{ind},1}^t,a_{\text{ind},2}^t, \cdots ,a_{\text{ind},M}^t)^\top,
\end{equation}
where
\begin{equation*}
    a_{\text{ind},m}^t = \begin{cases}
        0,& \hspace{-0.1cm}\text{if}~\epsilon^*(s_m^t) \ge 0 ~\text{and}~ \epsilon^*(s_m^t) \in {\mathcal{I}nd}^t;\\
        1,& \hspace{-0.1cm}\text{others},
    \end{cases}
\end{equation*}
and ${\mathcal{I}nd}^t$ consists of $K$ largest elements of  $\{\epsilon^*(s_1^t),\allowbreak\epsilon^*(s_2^t),\allowbreak\cdots,\allowbreak\epsilon^*(s_M^t)\}$.

\end{defi}


\section{Lower Bound and State-Independent Policies}\label{sec:V}
As discussed in Section \ref{sec:IIIA}, the optimal policy for the dynamic on-off switching problem is computationally prohibitive, especially when $M$ is large.
To evaluate the performance of the index policy, this section constructs lower and upper bounds as benchmarks.

\begin{thm}[Lower bound of the average cost]\label{thm_lower_bound}
A lower bound of the average cost in \eqref{eqA1} is given by
\begin{equation}\label{eqG0}
    L_B =\sum_{m=1}^M C_m^{(0)} + \sum_{m=1}^M \sum_{\ell>\gamma_m^L} \Delta_{2,m} (\ell) \cdot \Pr(\widetilde n_m = \ell),
\end{equation}
where
$$\Delta_{2,m} (\ell) \triangleq f\left[ {{\mathcal{P}}_\text{static}+( \ell+\overline{\lambda}_m T_s ) {\mathcal{P}}_d } \right] - f\left[ {(\ell+\overline{\lambda}_m T_s ) {\mathcal{P}}_e } \right],$$
$\ell \in \mathbb{N}$, $C_m^{(0)}$ is the anticipated cost of a time segment when $a_m^t=0$ and it is defined in \eqref{eqD5}, $\gamma_m^L = \frac{{\mathcal{P}}_\text{static}}{{\mathcal{P}}_e-{\mathcal{P}}_d} - \overline{\lambda}_m T_s$ is a threshold defined in Theorem~\ref{thm_greedy_structure}, and $\Pr(\widetilde n_m = \ell)$ can be calculated by \eqref{eqB1}.
\end{thm}

\begin{NewProof}
See Appendix~\ref{sec:AppD}.
\end{NewProof}

Next, we analyze the performance of two state-independent policies, i.e., the uniform policy and the round-robin policy, as upper bounds of the optimal policy.

With the uniform policy, we turn off $K$ gNBs uniformly at random in each time segment. Its performance is characterized in Proposition~\ref{prop_uniform_policy_performance}.

\begin{prop}[Performance of the uniform policy]\label{prop_uniform_policy_performance}
The long-term average cost of the uniform policy is given by
\begin{eqnarray}\label{eqH1}
    \overline{C}_{\text{uniform}} = 
    && \hspace{-0.5cm} \sum_{m = 1}^M \left[C_m^{(01)} \cdot (1-K/M) \cdot K/M\right. \\
    && \left. + {C_m^{(11)} \cdot (1-K/M)^2} + {C_m^{(0)} \cdot K/M} \right], \nonumber
\end{eqnarray}
where $C_m^{(01)}$, $C_m^{(11)}$ and $C_m^{(0)}$ are defined in \eqref{eqD5}. 
\end{prop}

\begin{NewProof}
See Appendix~\ref{sec:AppE}.
\end{NewProof}




\begin{figure}[t]
\centering
\includegraphics[width=1\columnwidth]{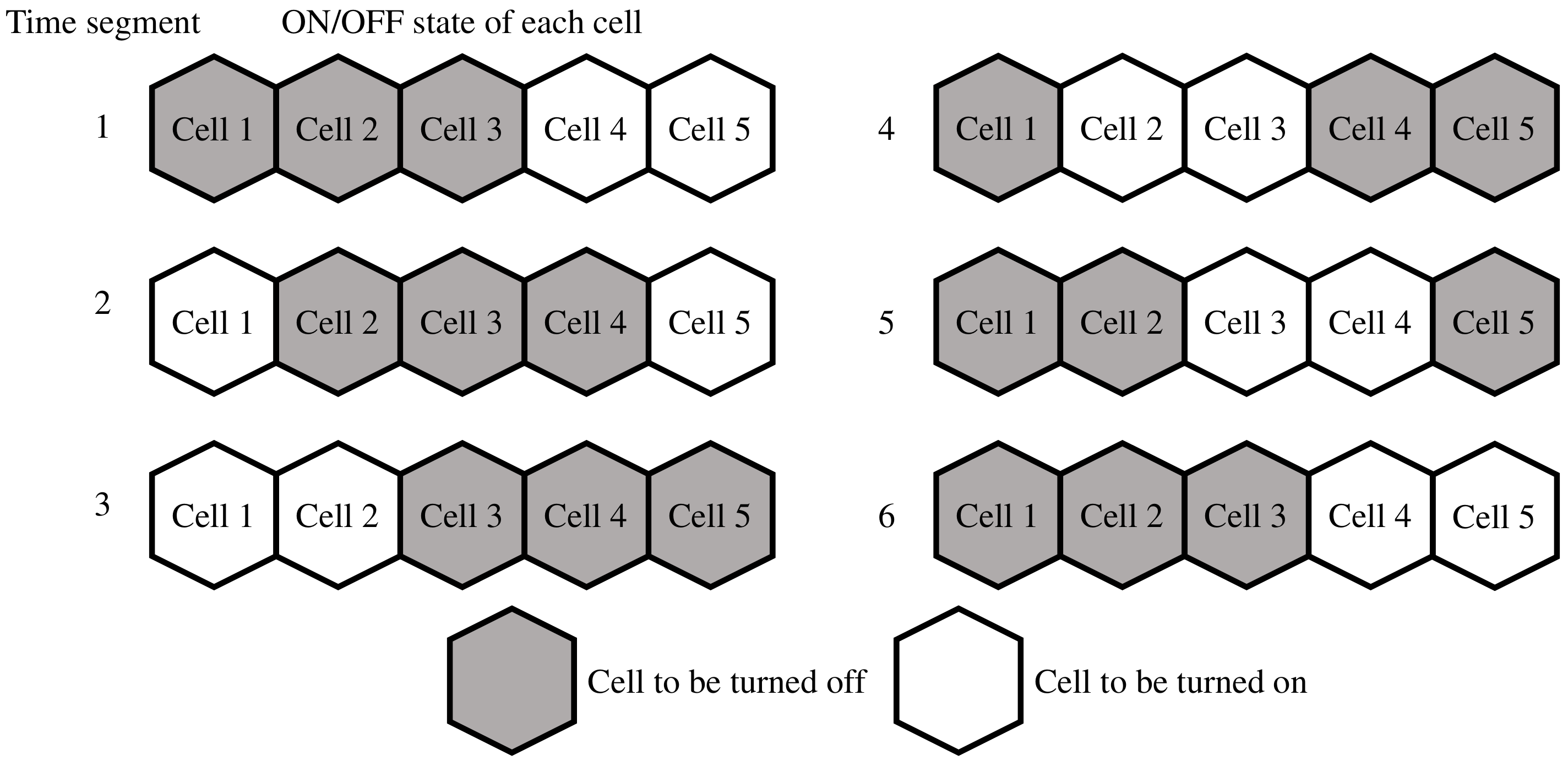}
\caption{An illustration of the round-robin policy.}
\label{fig:round-robin}
\end{figure}

The round-robin policy, on the other hand, turns off the gNBs in a deterministic order, an example of which is shown in Fig.~\ref{fig:round-robin}.
As can be seen, there are five 5G cells in the cluster and three gNBs will be turned off in each time segment.
With the round-robin policy, each gNB will be turned off in $K$ consecutive time segments and turned on in the following $M-K$ consecutive time segments. Its performance is characterized in Proposition~\ref{prop_robin_policy_performance}.
\begin{prop}[Performance of the round-robin policy]\label{prop_robin_policy_performance}
The long-term average cost of the round-robin policy is given by
\begin{equation}\label{eqH5}
\overline{C}_{\text{round}} = \begin{cases}
        \sum_{m = 1}^M C_m^{(11)}, & K=0; \\
        \sum_{m = 1}^M C_m^{(0)}, & K=M; \\
        \frac{1}{M} \sum_{m = 1}^M \left[C_m^{(0)} K + C_m^{(01)}\right. & \\
        \hspace{0.5cm} \left. + C_m^{(11)} (M-K-1) \right],&0<K<M,
    \end{cases}
\end{equation}
where $C_m^{(01)}$, $C_m^{(11)}$ and $C_m^{(0)}$ are as defined in \eqref{eqD5}.
\end{prop}

\begin{NewProof}
See Appendix~\ref{sec:AppF}.
\end{NewProof}

    

As can be seen, the long-term average cost of the round-robin policy is linear with $K$ when $0<K<M$, and is discontinuous when $K=0$ and $K=M$.

In the following, we compare the performance of these two state-independent policies.
When $K=0$ (or $K=M$), their performances are the same. When $0<K<M$, their difference is
\begin{eqnarray}\label{eqH10}
    \overline{C}_{\text{diff}} \hspace{-0.3cm}&\triangleq&\hspace{-0.3cm} \overline{C}_{\text{round}} - \overline{C}_{\text{uniform}} \\
    \hspace{-0.3cm}&=&\hspace{-0.3cm} \frac{K^2-MK+M}{M^2} \sum_{m = 1}^M { C_m^{(01)}-C_m^{(11)} } \nonumber \\
    \hspace{-0.3cm}&=&\hspace{-0.3cm} \frac{K^2-MK+M}{M^2} C_{\text{diff}}, \nonumber
\end{eqnarray}
where $C_{\text{diff}} \triangleq \sum_{m = 1}^M { C_m^{(01)}-C_m^{(11)} }$ is a constant and we have $C_{\text{diff}} > 0$ when $\mathcal{P}_\text{switch} > 0$.

As shown in \eqref{eqH10}, the difference $\overline{C}_{\text{diff}}$ is quadratic in $K$, the maximum $\overline{C}_{\text{diff-max}}$ is achieved when $K=1$ and $K=M-1$, and the minimum $\text{or}~\overline{C}_{\text{diff-min}}$ is achieved when $K=M/2$ (for even $M$) or $K=(M \pm 1)/2$ (for odd $M$). Specifically,
\begin{eqnarray*}\label{eqH11}
    \overline{C}_{\text{diff-max}} \hspace{-0.3cm}&= &\hspace{-0.3cm} \frac{1}{M^2} C_{\text{diff}}, \hspace{1.7cm} K=1~ \text{or}~M-1; \\
    \overline{C}_{\text{diff-min}} \hspace{-0.3cm}&=&\hspace{-0.3cm}
    \begin{cases}
        \frac{4-M}{4M} C_{\text{diff}},& K=M/2; \\
        \frac{-M^2+4M+1}{4} C_{\text{diff}},& K=(M \pm 1)/2.
    \end{cases}
\end{eqnarray*}

As a result, the uniform policy is strictly better than the round-robin policy when
\begin{enumerate}
\item $K=1$ or $M-1$, in which case $\overline{C}_{\text{diff-max}} > 0$;
\item $M<4$, in which case $\overline{C}_{\text{diff-min}} > 0$.
\end{enumerate}


\section{Numerical and Simulation Results}\label{sec:VI}
This section presents numerical and simulation results to evaluate various policies analyzed in this paper, i.e., the optimal policy, the greedy policy, the index policy, and the state-independent policies. In particular, we measure the performance of a policy by the gap between the long-term average cost of this policy and the lower bound given in Theorem~\ref{thm_lower_bound}. That is:
\begin{equation}\label{eqI1}
    \Delta_{\text{policy}} \triangleq \frac{{\overline C}_{\text{policy}}-L_B}{L_B} \times 100\%
\end{equation}
where the lower bound $L_B$ is defined in \eqref{eqG0} and ${\overline C}_{\text{policy}}$ denotes the long-term average cost of an evaluated policy over the infinite-time horizon.

The parameter settings are presented in Table~\ref{tab:simulation_parameters} unless specified otherwise. Specifically, as \cite{7445140,6736750,9435313}, we set the average power consumption of a gNB for serving a single user to ${\mathcal{P}}_d=1$W; the static power consumption of a gNB ${\mathcal{P}}_\text{static}=85$W; the switching power consumption of a gNB ${\mathcal{P}}_\text{switch}=50$W.
The average power consumption of the ng-eNB for serving a single user ${\mathcal{P}}_e$ is set to 5W.
The duration of a time segment $T_s$ is set to 1800s.
Users' arrival to a cell follows a mixed Poisson process with a set of parameters $\Lambda=\{0.005,0.01,0.015,0.02\}$.
We consider five sets of sampling probabilities $\Pr(\Lambda)$, as listed in Table~\ref{tab:simulation_parameters}.
The mean service time of a user is set to $1/{\mu}=500$s \cite{8866747}.
We consider three kinds of cost
functions $f$ in this paper: the quadratic function $f(x)=x^2$; the piecewise linear function
$$f(x)=\begin{cases}
0.5x,& \text{if}~x \le 100;\\
x-50,& \text{if}~100 < x \le 150;\\
1.5x-125,& \text{if}~x >150;
\end{cases}$$
and the linear function $f(x)=x$.

\begin{table}[t]
\caption{Parameter settings.}
\centering
\begin{tabular}{cccc}
\toprule
\textbf{Physical Quantities}                                 & \textbf{Symbols} & \textbf{Values} & \textbf{Units} \\ \midrule
Average power/gNB/user                               & ${\mathcal{P}}_d$ &  1      &  W     \\
Average power of macro BS/user                          & ${\mathcal{P}}_e$ &  5      & W \\
Static power/gNB                                     & ${\mathcal{P}}_\text{static}$ & 85       & W      \\
Switching power/gNB                                  & ${\mathcal{P}}_\text{switch}$ & 40       & W      \\ \midrule
Time segment duration                                & $T_s$          &  1800      &  s     \\
Mean service time/user                               & $1/\mu$          & 500  & s     \\
User arrival rate                                    & $\Lambda$ & 
$\begin{Bmatrix}
\begin{smallmatrix}
0.005 & 0.01 \\
0.015 & 0.02
\end{smallmatrix}
\end{Bmatrix}$ & /s      \\ \midrule
\multirow{5}{*}{Sampling probabilities $\Pr(\Lambda)$} &  \multicolumn{3}{c}{Set1:~$\{0,1,0,0\}$}        \\
                                                     &  \multicolumn{3}{c}{Set2:~$\{0.5,0,0.5,0\}$}        \\
                                                     & \multicolumn{3}{c}{Set3:~$\{2/3,0,0,1/3\}$}        \\ 
                                                     &  \multicolumn{3}{c}{Set4:~$\{0.3,0.4,0.3,0\}$}        \\ 
                                                     &  \multicolumn{3}{c}{Set5:~$\{0.6,0,0.2,0.2\}$}        \\ 
\bottomrule
\end{tabular}
\label{tab:simulation_parameters}
\end{table}

\begin{figure}[t]
\centering
\includegraphics[width=0.8\columnwidth]{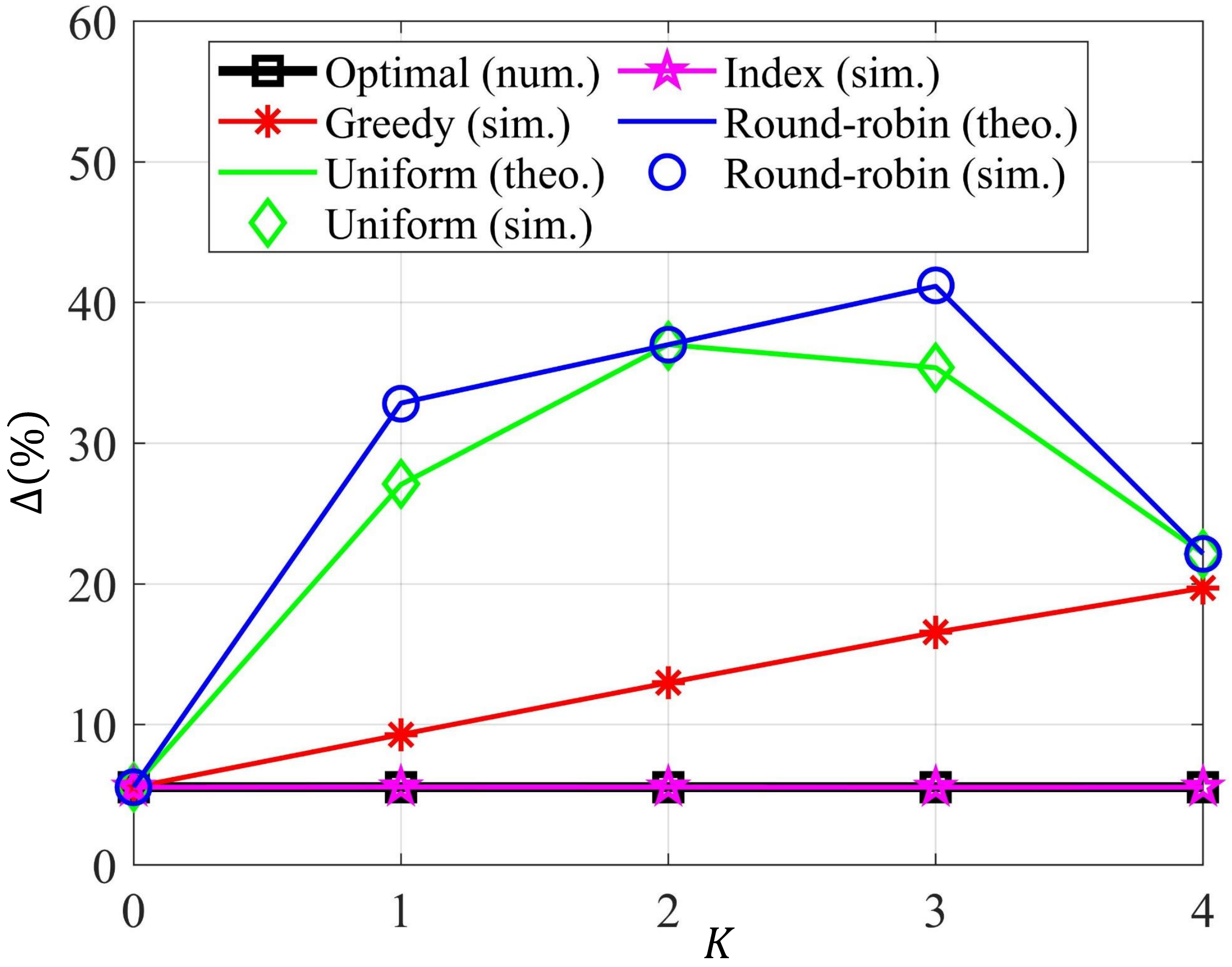}
\caption{Performance of different switching policies versus $K$, where $M=4$, $\Pr(\Lambda)=\{2/3,0,0,1/3\}$, and $f(x)=x^2$.}
\label{fig:result_K_M4}
\end{figure}

According to \eqref{eqC4}, the optimal policy is computationally prohibited for large $M$. Thus, in the first simulation, we consider a small cluster with $M=4$ 5G cells and evaluate the performance of various policies benchmarked against the optimal policy.
The simulation results are presented in Fig.~\ref{fig:result_K_M4}, where we plot $\Delta_{\text{policy}}$ as a function of $K$.
As can be seen, the index policy achieves the same performance as the optimal policy and outperforms the other policies.
The greedy policy is better than state-independent policies.
The cost of the uniform policy $\Delta_{\text{uniform}}$ increases quadratically in $K$, as predicted in \eqref{eqH1}.
The cost of the round-robin policy $\Delta_{\text{round-robin}}$, on the other hand, increases linearly when $1 \le K \le M-1$, as predicted in \eqref{eqH5}.

\begin{figure}[t]
\centering
\includegraphics[width=1\columnwidth]{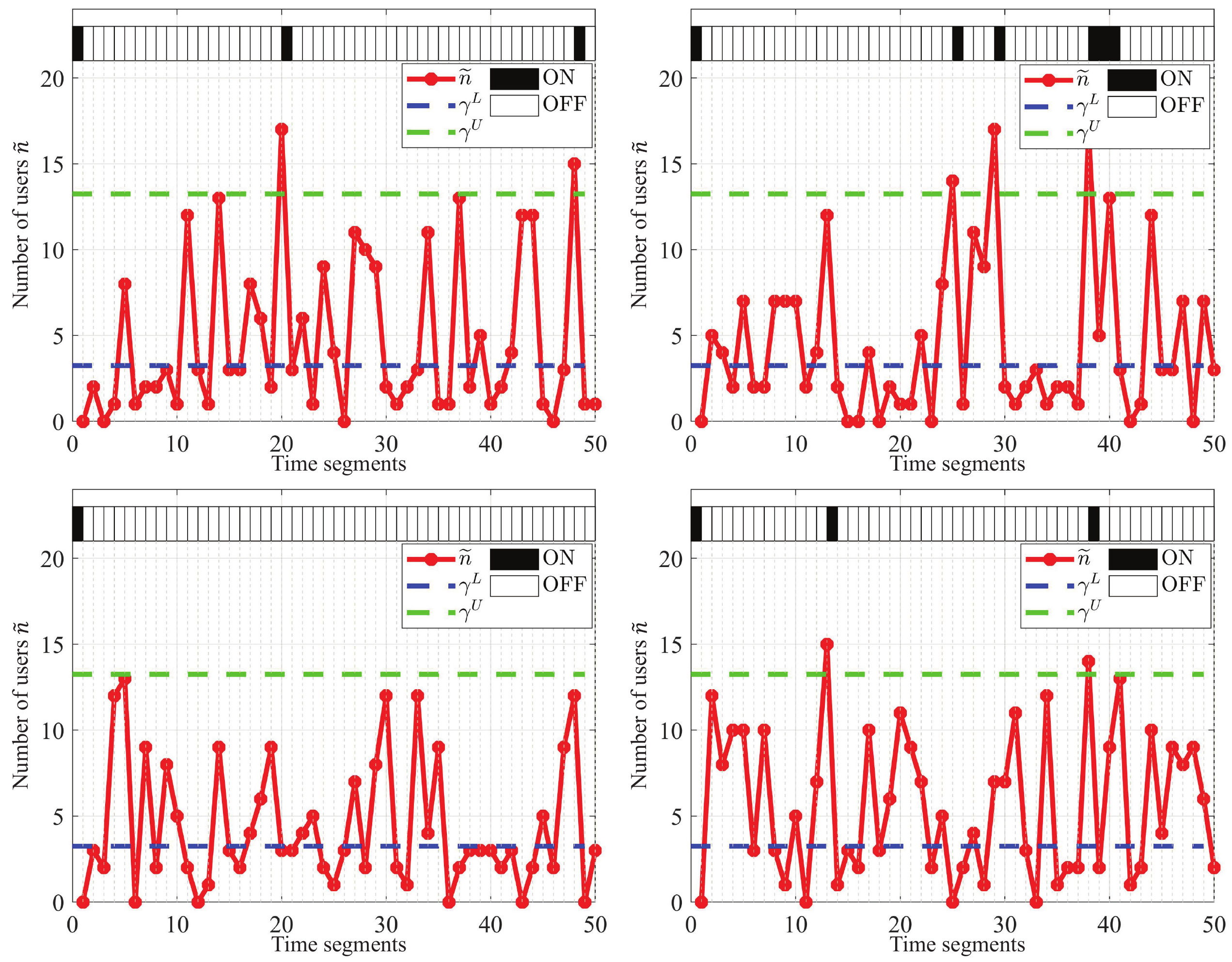}
\caption{The dual-threshold structure of the greedy policy, where $M=K=4$, $\Pr(\Lambda)=\{2/3,0,0,1/3\}$, and $f(x)=x^2$.}
\label{fig:result_greedy_on_off}
\end{figure}

Recall from Theorem \ref{thm_greedy_structure} that the greedy policy exhibits a dual-threshold structure when $K=M$.
This is verified in  Fig.~\ref{fig:result_greedy_on_off}, where we set $K=M=4$.
For each 5G cell, we plot the number of residual users at the beginning of a time segment and the corresponding ON/OFF actions of the greedy policy.
As shown, for each cell, the greedy policy presents a dual-threshold structure: when the number of residual users $\widetilde{n}$ at the beginning of a time segment is smaller than $\gamma_m^L$, the gNB will be turned off; when $\widetilde{n}$ is larger than $\gamma_m^U$, the gNB will be turned on; when $\gamma_m^L \leq \widetilde{n}\leq \gamma_m^U$, the ON/OFF state of a gNB remains unchanged.

\begin{figure}[t]
\centering
\includegraphics[width=0.8\columnwidth]{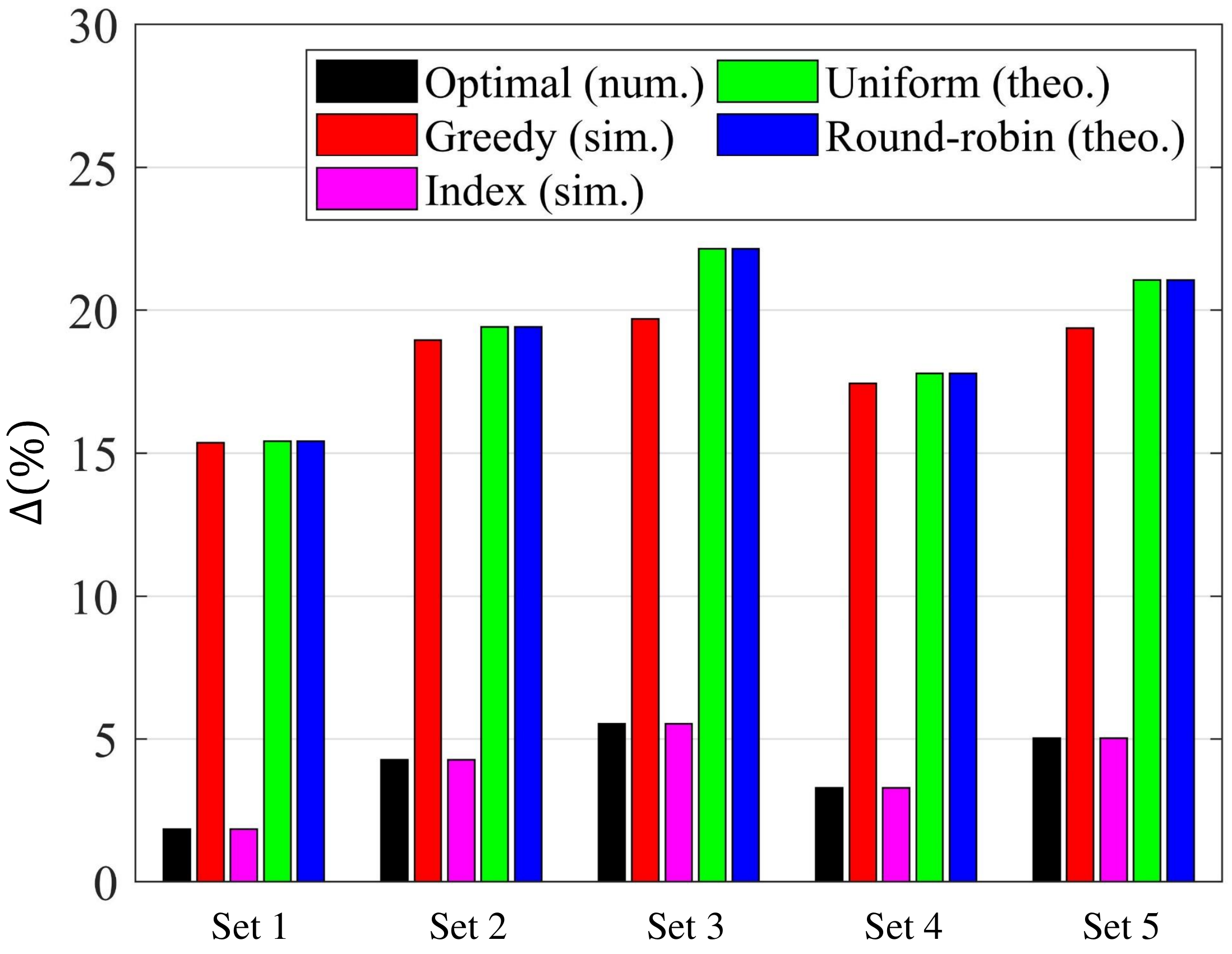}
\caption{Impact of user arrival distributions $\Pr(\Lambda)$ on the performance of various switching policies, where $M=K=4$ and $f(x)=x^2$.}
\label{fig:result_user_coming_model}
\end{figure}

In the second simulation, we evaluate the impact of user arrival distributions on the performance of various switching policies.
Specifically, we extend the simulation in Fig.~\ref{fig:result_K_M4} to different user arrival distributions considering the five sets of $\Pr(\Lambda)$ listed in Table~ \ref{tab:simulation_parameters} (note that the mean user arrival rates are the same under the five sets of distributions).
The simulation results are presented in Fig.~\ref{fig:result_user_coming_model}. As shown, the index policy always achieves close-to-optimal performance under different sets of $\Pr(\Lambda)$.

\begin{figure}[t]
\centering
\includegraphics[width=0.8\columnwidth]{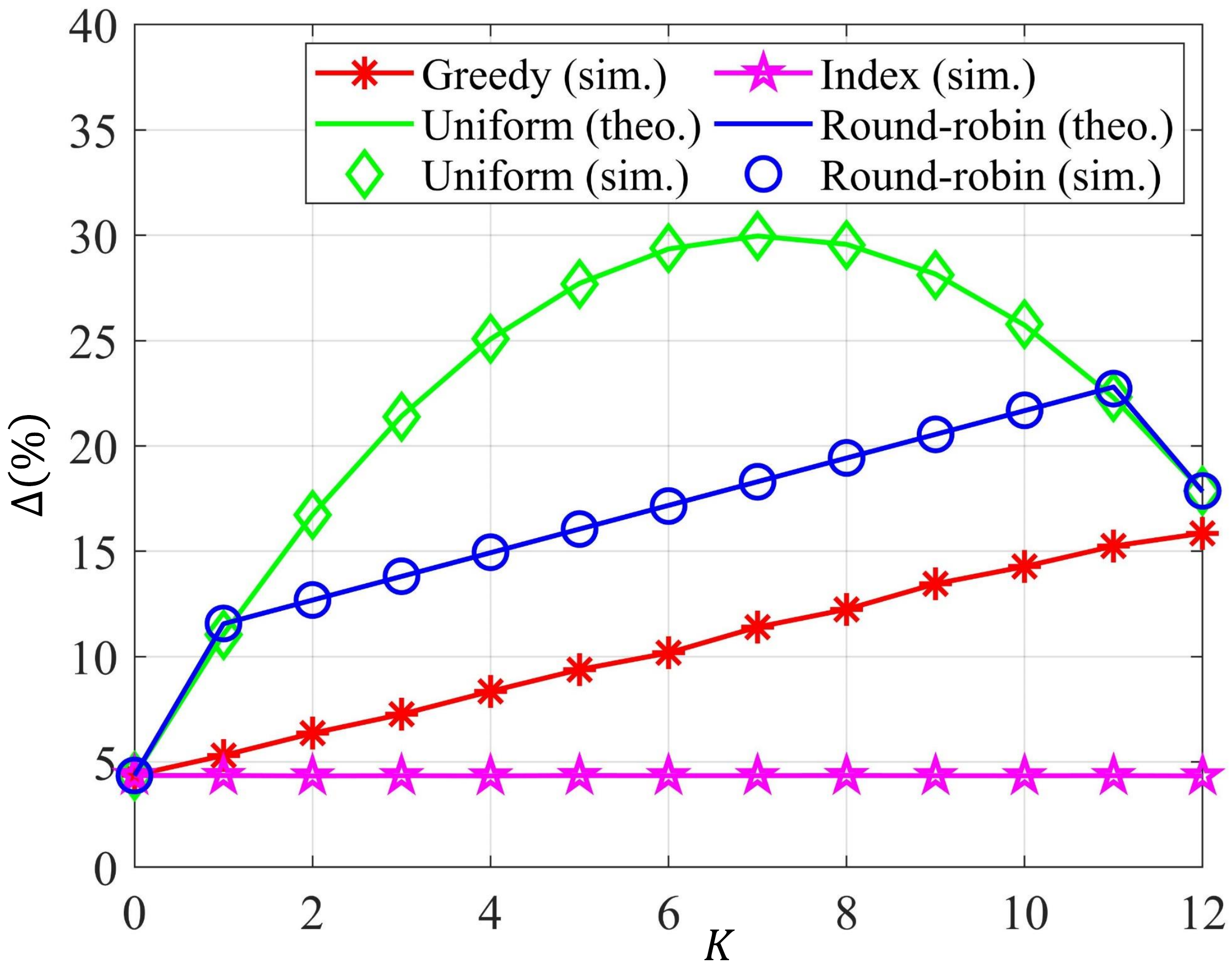}
\caption{Performance of different switching policies versus $K$, where $M=12$, $\Pr(\Lambda)=\{2/3,0,0,1/3\}$, and $f(x)$ is a piecewise linear function.}
\label{fig:result_K_M12_fpiecewise}
\end{figure}

\begin{figure}[t]
\centering
\includegraphics[width=0.8\columnwidth]{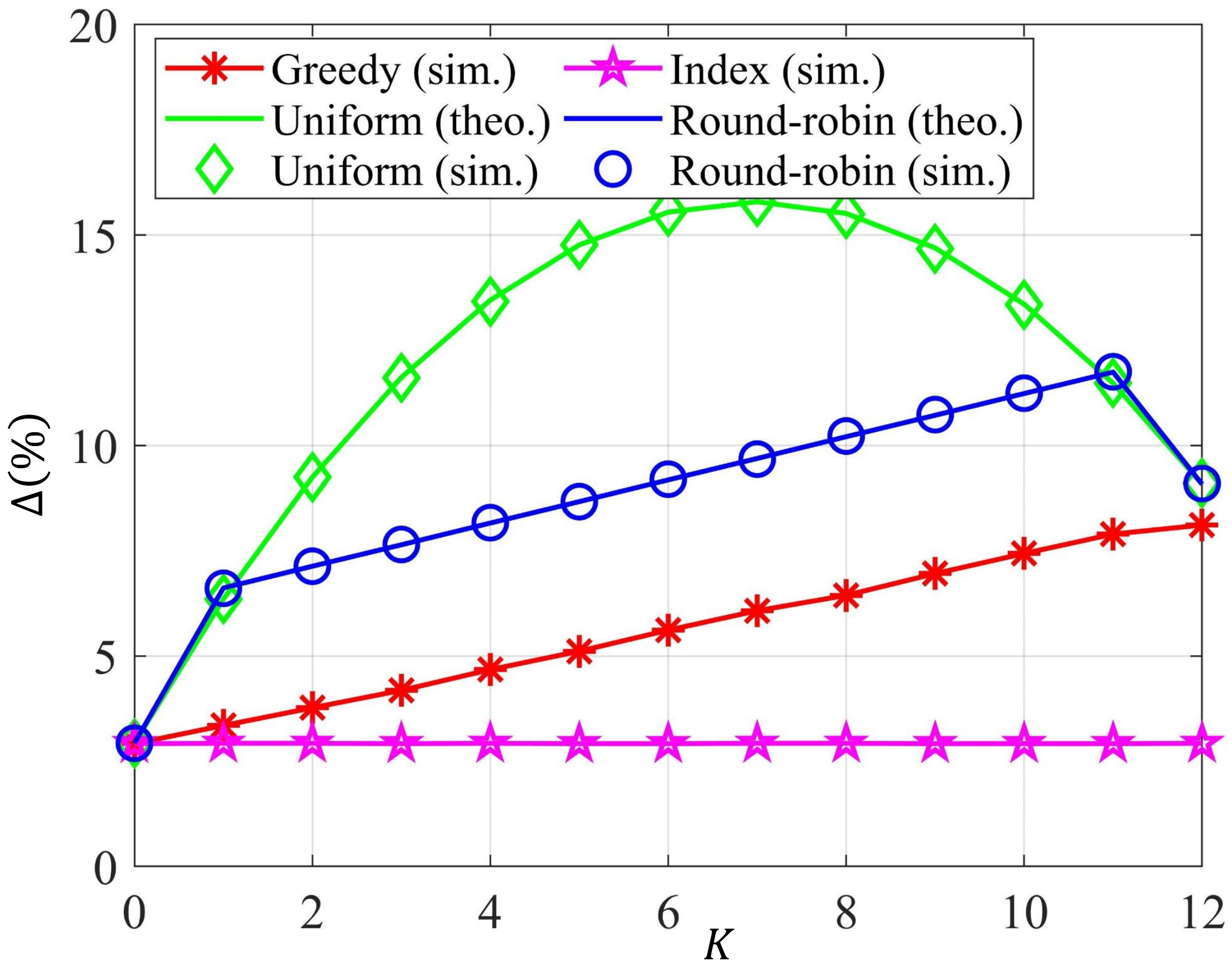}
\caption{Performance of different switching policies versus $K$, where $M=12$, $\Pr(\Lambda)=\{2/3,0,0,1/3\}$, and $f(x)=x$.}
\label{fig:result_K_M12_fx}
\end{figure}

In the third simulation, we consider a larger cluster with $M=12$ cells, in which case the performance of the optimal policy is no longer available, and we shall consider different cost functions $f(x)$. 
Fig.~\ref{fig:result_K_M12_fpiecewise} and \ref{fig:result_K_M12_fx} present the performance of various policies versus $K$ with the piecewise linear and linear cost functions, respectively.
We have similar observations as that from Fig.~\ref{fig:result_K_M4}, the index policy achieves the best performance among all switching policies and the properties of other policies match our predictions.

\begin{figure}[t]
\centering
\includegraphics[width=0.8\columnwidth]{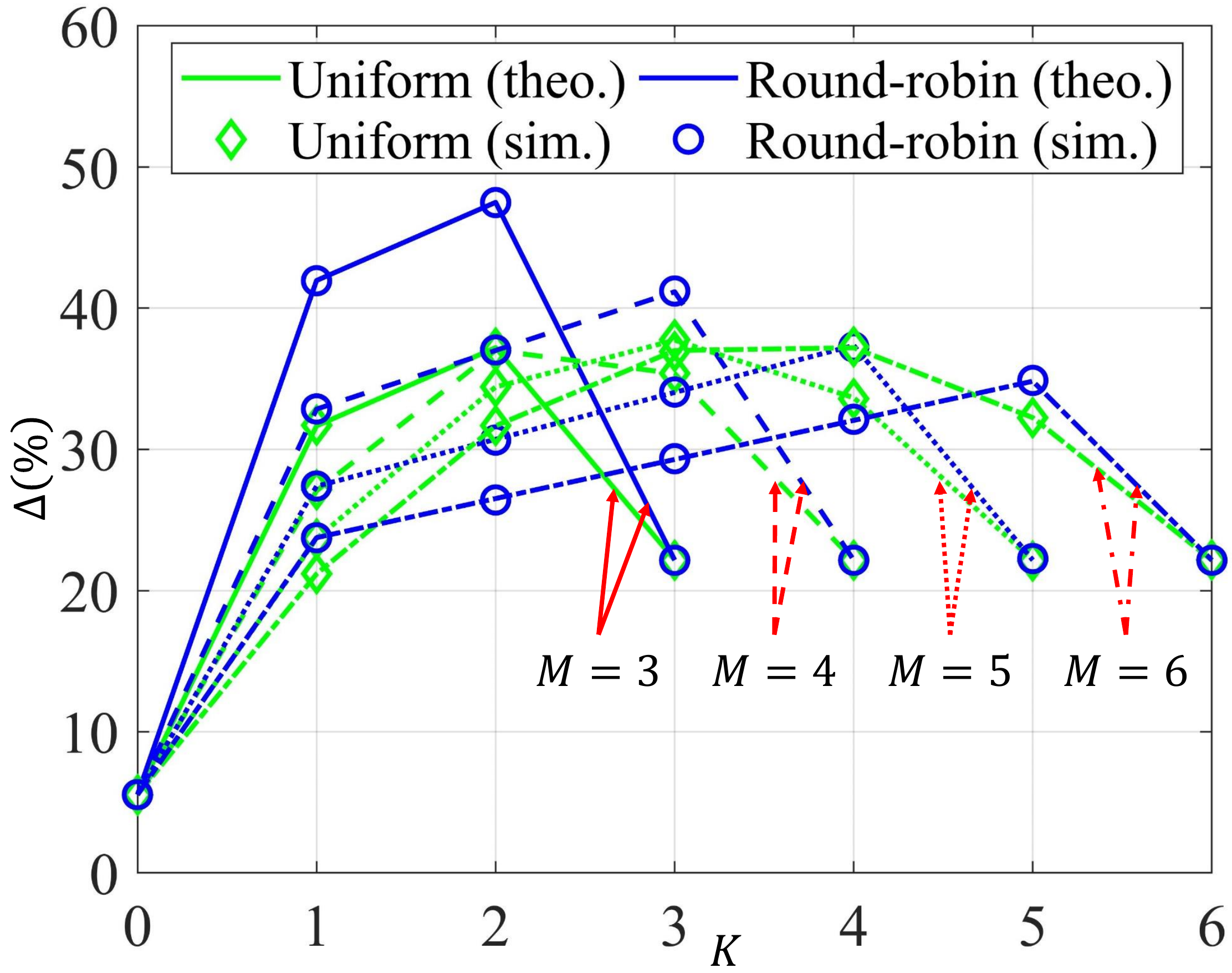}
\caption{Comparison of the two state-independent policies, where $\Pr(\Lambda)=\{2/3,0,0,1/3\}$, and $f(x)=x^2$.}
\label{fig:result_compare_stateindependent}
\end{figure}

Next, we compare the performance of the uniform and round-robin policies in more detail to verify our analysis in Section~\ref{sec:V}. As can be seen from Fig.~\ref{fig:result_compare_stateindependent}, the uniform policy is strictly better than the round-robin policy when $M<4$ and $K=1$ or $K=M-1$. In other cases, the round-robin policy can be better than the uniform policy.

\begin{figure}[t]
\centering
\includegraphics[width=0.8\columnwidth]{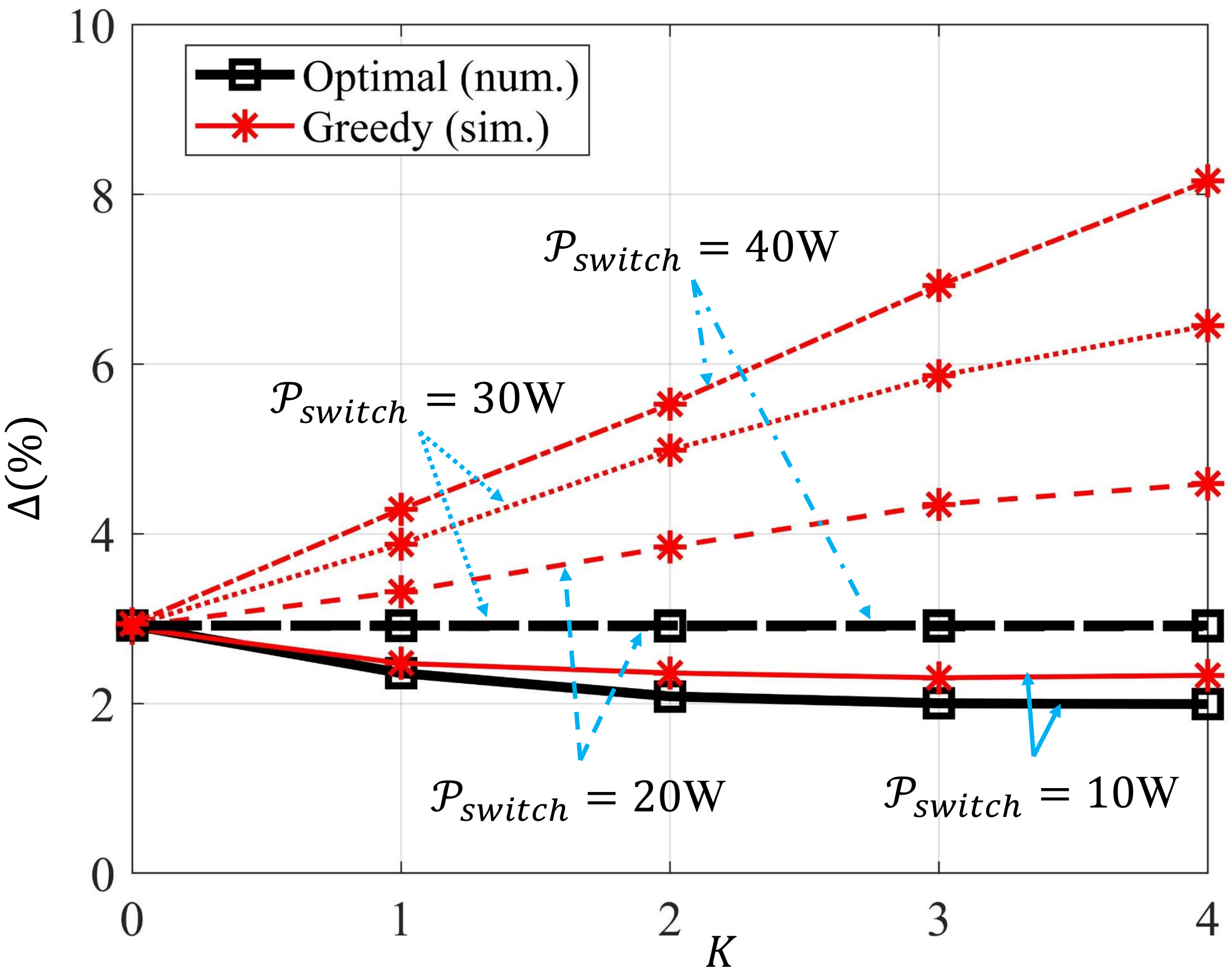}
\caption{Impact of the switching power consumption on the greedy policy and the optimal policy, where $M=4$, ${\mathcal{P}}_\text{switch}=20$W, $\Pr(\Lambda)=\{2/3,0,0,1/3\}$, and $f(x)=x$.}
\label{fig:result_K_M4_Pswi20}
\end{figure}

\begin{figure}[t]
\centering
\includegraphics[width=0.8\columnwidth]{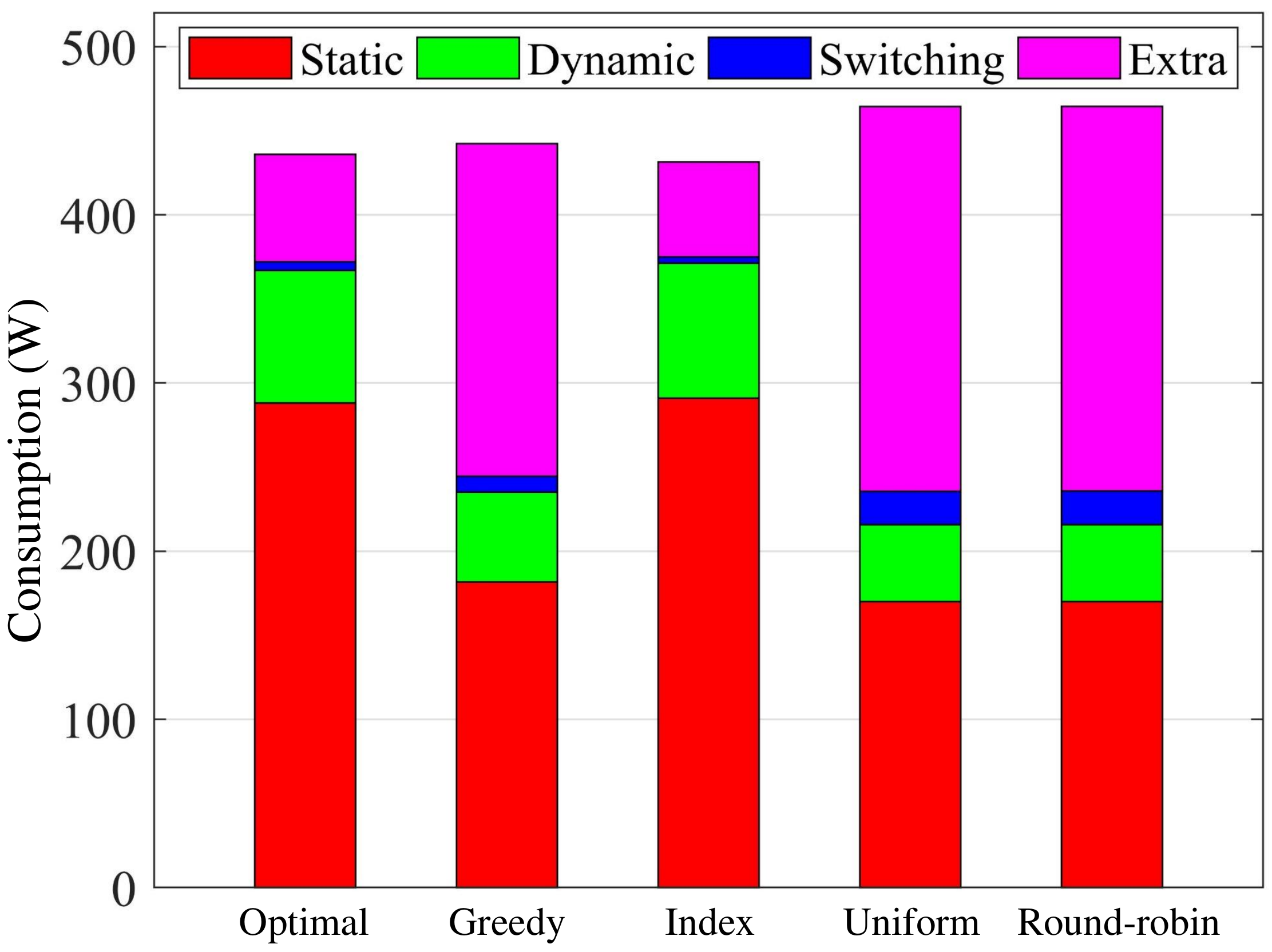}
\caption{The compositions of the power consumption for different policies, where $M=4$, ${\mathcal{P}}_\text{switch}=20$W, $K=2$, $\Pr(\Lambda)=\{2/3,0,0,1/3\}$, and $f(x)=x$.}
\label{fig:result_composition}
\end{figure}

It is worth noting that the optimality of the greedy policy is determined by the switching cost ${\mathcal{P}}_\text{switch}$. In the extreme case where the switching cost is zero and frequent switching is allowed, the greedy policy is optimal. In Fig.~\ref{fig:result_K_M4_Pswi20}, we evaluate the impact of the switching cost by reducing the power consumption of switching from ${\mathcal{P}}_\text{switch}=40$W to ${\mathcal{P}}_\text{switch}=10$W. In the simulation, we consider a linear cost function $f(x)=x$ and $M=4$. As shown, with the decrease in the switching cost, the gap between the greedy policy and the optimal policy reduces.


For the case of $K=2$ and ${\mathcal{P}}_\text{switch}=20$W in Fig.~\ref{fig:result_K_M4_Pswi20}, we further analyze the compositions of power consumption when operated with different policies. As can be seen from Fig.~\ref{fig:result_composition}, the index policy has almost the same composition of power consumption as the optimal policy, verifying that the index policy is close-to-optimal.
The compositions of the greedy policy, on the other hand, are much more different.
In general, the greedy policy is more inclined to turn off the gNBs compared with the optimal policy, and hence, incurs more power consumption of the ng-eNB and less power consumption of the gNBs.
A piece of theoretical evidence is given in Proposition~\ref{prop_greedy_optimal}, where we have proven that the number of ``ON'' time segments with the greedy policy is no greater than that with the optimal policy when $K=M$.
In addition, the greedy policy incurs more switching cost compared with the optimal policy.

\section{Conclusion}\label{sec:Conclusion}
To achieve energy-conserving 5G RAN, this paper put forth a dynamic on-off switching paradigm for gNB sleep control by taking the evolvements of users and their traffic demands into account.
Formulating the dynamic sleep control for a cluster of gNBs as an MDP, we characterized the optimal policy for the MDP that minimizes the long-term average cost, where cost is a non-decreasing function of system energy expenditure.
The optimal policy, however, is computationally demanding and is available only when the number of gNBs is small.

To meet this challenge, we proposed a greedy policy and an index policy and analyzed their performances benchmarked against the optimal policy and state-independent policies.
When making on-off switching decisions, the greedy policy focuses only on the immediate effect of the decisions while omitting their long-term impacts. We proved the dual-threshold structure of the greedy policy when there is no constraint on the number of gNBs that can be turned off, and analyzed its connections with the optimal policy.
On the other hand, the index policy assigns an index to each gNB as a measurement of how one is willing to pay to turn off the gNB.
The gNBs with relatively larger indexes are then turned off when making switching decisions.
To develop the index policy, we decoupled the original MDP, proved the indexibility of the decoupled MDP, and proposed an algorithm to compute the index.
Although heuristic, the index policy exhibits close-to-optimal performance and outperforms the greedy policy and state-independent policies. Furthermore, it is much more computationally efficient than the optimal policy.

To summarize, our study validated the effectiveness of gNB sleep control in achieving an energy-efficient 5G RAN.
Under the established dynamic on-off switching paradigm, we demonstrated that the proposed switching policies can reduce the system energy expenditure by a large margin, providing useful operational insights for practical 5G RAN.

\appendices

\section{Proof of Proposition~\ref{prop_residual_users_distribution}}\label{sec:AppA}
Considering the $(t-1)$-th time segment, the $m$-th cell serves two sets of users: the newly arrived users as well as the residual users from the $(t-2)$-th time segment. For simplicity, we refer to the two sets as $\mathcal{G}_n$ and $\mathcal{G}_r$,  respectively. At the end of the $(t-1)$-th time segment, the number of residual users in the cell can be written as
\begin{equation*}
    \widetilde n_m^t = \ell_m^{t-1} + \widetilde{\ell}_m^{t-1},
\end{equation*}
where $\ell_m^{t-1}$ and $\widetilde{\ell}_m^{t-1}$ are the number of residual users from $\mathcal{G}_n$ and $\mathcal{G}_r$, respectively.

First, $\ell_m^{t-1}$ can be written as
\begin{equation*}
    \ell_m^{t-1} = \sum_{n=1}^{n_m^{t-1}} I(T_s,\xi_n,\tau_n),
\end{equation*}
where $\tau_n$ and $\xi_n$ denote the arrival epoch and staying time of the $n$-th user in $\mathcal{G}_n$, respectively; $n_m^{t-1}$ follows the mixed Poisson distribution and $\xi_n$ follows the exponential distribution as described in Definition~\ref{defi_user_process}; The function $I(\cdot)$ indicates whether the $n$-th user is still in the cell at the end of the $(t-1)$-th time segment. That is,
\begin{equation*}
    I(T_s,\xi_n,\tau_n) = \begin{cases}
        1, & \text{if}~T_s-\tau_n \le \xi_n;\\
        0, & \text{otherwise}.
    \end{cases}
\end{equation*}

For a specific parameter $\lambda_j \in \Lambda$ of the mixed Poisson process, $\ell_m^{t-1}$ is a filtered Poisson process \cite{snyder2012random}. The characteristic function of $\ell_m^{t-1}$ can be written as
\begin{eqnarray}\label{eqAppA1}
    \phi_{\ell_m^{t-1}} \hspace{-0.3cm}&\triangleq&\hspace{-0.3cm} \mathbb{E} [e^{j \omega \ell_m^{t-1}}] = \mathbb{E} \left[\exp \left( {j \omega \sum_{n=1}^{n_m^{t-1}} I(T_s,\xi_n,\tau_n) } \right) \right] \nonumber \\
    &=&\hspace{-0.3cm} \mathbb{E}_{n_m^{t-1},\{\tau_n\}} \left\{ \mathbb{E}_{\{\xi_n\}} \left[ { \exp \left( {j \omega \sum_{n=1}^{n^*} I(T_s,\xi_n,\tau_n) } \right) } \right] \right. \nonumber \\
    &&\hspace{1.5cm} \left. | {n_m^{t-1}=n^*,\{\tau_n\}} \right\} .
\end{eqnarray}

Let $B_n(T_s,\tau_n) \triangleq \mathbb{E}_{\xi_n} \left[ { \exp \left( {j \omega I(T_s,\xi_n,\tau_n) } \right) } \right]$, \eqref{eqAppA1} can be refined as
\begin{eqnarray}\label{eqAppA2}
    \phi_{\ell_m^{t-1}} \hspace{-0.3cm}&=&\hspace{-0.3cm} \mathbb{E}_{n_m^{t-1},\{\tau_n\}} \left[ \prod_{n=1}^{n^*} B_n(T_s,\tau_n) \left| n_m^{t-1}=n^*,\{\tau_n\} \right. \right] \nonumber\\
    &=&\hspace{-0.3cm} \mathbb{E}_{n_m^{t-1}} \left[ \mathbb{E}_{\{\tau_n\}} \left( \prod_{n=1}^{n^*} B_n(T_s,\tau_n) \right) \left| n_m^{t-1}=n^* \right. \right] \nonumber\\
    &\overset{(a)}{=}&\hspace{-0.3cm} \mathbb{E}_{n_m^{t-1}} \left[ \int_0^{T_s} \int_0^{\tau_{n^*}} \cdots \int_0^{\tau_2} \prod_{n=1}^{n^*} B_n(T_s,\tau_n) \frac{n!}{T_s^{n^*}} \right. \nonumber\\
    &&\hspace{1.0cm} \left. d\tau_1 \cdots d\tau_{n^*-1} d\tau_{n^*} \left| n_m^{t-1}=n^* \right. \right] \nonumber\\
    &=&\hspace{-0.3cm} \mathbb{E}_{n_m^{t-1}} \left[ \frac{1}{T_s^{n^*}} \int_0^{T_s} \int_0^{T_s} \cdots \int_0^{T_s} \prod_{n=1}^{n^*} B_n(T_s,\tau_n) \right. \nonumber\\
    &&\hspace{1.0cm} \left. d\tau_1 \cdots d\tau_{n^*-1} d\tau_{n^*} \left| n_m^{t-1}=n^* \right. \right] \nonumber\\
    &\overset{(b)}{=}&\hspace{-0.3cm} \mathbb{E}_{n_m^{t-1}} \left[ \left( \frac{1}{T_s} \int_0^{T_s} B(T_s,\tau) d\tau \right)^{n_m^{t-1}} \right] \nonumber\\
    &\overset{(c)}{=}&\hspace{-0.3cm} G_{n_m^{t-1}} \left( \frac{1}{T_s} \int_0^{T_s} B(T_s,\tau) d\tau \right),
\end{eqnarray}
where $(a)$ follows because the arrival epochs $\tau_n~(n=1,\allowbreak 2,\allowbreak \cdots,\allowbreak n_m^{t-1})$ given $n_m^{t-1}$ have a distribution as the order statistics sampled from a uniform distribution \cite{snyder2012random}; $(b)$ follows because all $\tau_n$ has the same distribution such that the subscript $n$ can be omitted; $(c)$ follows from the definition of the moment generating function $G_X(z) \triangleq \mathbb{E}(z^X)$.

Note that the random variable $n_m^{t-1}$ follows the Poisson distribution, \eqref{eqAppA2} can be further written as
\begin{equation}\label{eqAppA3}
    \phi_{\ell_m^{t-1}} = \exp \left[\lambda_j T_s \left( \frac{1}{T_s} \int_0^{T_s} B(T_s,\tau) d\tau -1 \right) \right].
\end{equation}
In particular, $B(T_s,\tau)$ is given by
\begin{eqnarray}\label{eqAppA4}
    B(T_s,\tau) \hspace{-0.3cm}&=&\hspace{-0.3cm} \mathbb{E}_{\xi} [ \exp (j \omega I(T_s,\xi,\tau) )] \nonumber\\
    &=&\hspace{-0.3cm} \Pr[I(T_s,\xi,\tau)=0] + e^{j \omega} \cdot \Pr[I(T_s,\xi,\tau)=1] \nonumber \\
    &=&\hspace{-0.3cm} \Pr[I(T_s,\xi,\tau)=1] \cdot (e^{j \omega}-1) +1 \nonumber \\
    &=&\hspace{-0.3cm} \left( \int_{T_s-\tau}^{\infty} f_{\xi}(\xi) d\xi \right) \cdot (e^{j \omega}-1) +1 \nonumber \\
    &\overset{(a)}{=}&\hspace{-0.3cm} e^{-\mu_m (T_s-\tau)} \cdot (e^{j \omega}-1) +1, 
\end{eqnarray}
where $(a)$ holds because the staying time $\xi$ follows the exponential distribution with parameter $\mu_m$. By substituting \eqref{eqAppA4} into \eqref{eqAppA3}, we have
\begin{eqnarray*}
    \phi_{\ell_m^{t-1}} \hspace{-0.3cm}&=&\hspace{-0.3cm} \exp \left( \frac{\lambda_j}{\mu_m} (1-e^{-\mu_m T_s})(e^{j \omega}-1) \right) \nonumber\\
    &=&\hspace{-0.3cm} \exp[\widetilde{\lambda}_{m,j} (e^{j \omega}-1)],
\end{eqnarray*}
where $\widetilde{\lambda}_{m,j} = \frac{\lambda_j}{\mu_m}(1-e^{-\mu_m T_s})$ is a constant. By Taylor series expansion, we have
\begin{equation}\label{eqAppA5}
    \phi_{\ell_m^{t-1}} = e^{-\widetilde{\lambda}_{m,j}} e^{\widetilde{\lambda}_{m,j} \cdot e^{j \omega}} = e^{-\widetilde{\lambda}_{m,j}} \sum_{\ell=0}^{\infty} \frac{\left( \widetilde{\lambda}_{m,j} \right)^{\ell} \cdot e^{j \omega \ell}}{\ell !}.
\end{equation}

Note that the characteristic function $\phi_{\ell_m^{t-1}}$ in \eqref{eqAppA1} can also be written as
\begin{equation}\label{eqAppA6}
    \phi_{\ell_m^{t-1}} = \mathbb{E} [e^{j \omega \ell_m^{t-1}}] = \sum_{\ell=0}^{\infty} \Pr(\ell_m^{t-1}=\ell) \cdot e^{j \omega \ell}.
\end{equation}

Comparing the coefficients of each items in \eqref{eqAppA5} and \eqref{eqAppA6} yields
\begin{equation*}
    \Pr(\ell_m^{t-1}=\ell) = \frac{\left( \widetilde{\lambda}_{m,j} \right)^{\ell} }{\ell !} e^{-\widetilde{\lambda}_{m,j}}.
\end{equation*}

For the mixed Poisson process, $\lambda_j$ is sampled from $\Lambda$ with probability $p_{m,j}$. Thus, the probability distribution of $\ell_m^{t-1}$ is given by
\begin{eqnarray*}
    \Pr(\ell_m^{t-1}=\ell) \hspace{-0.3cm}&=&\hspace{-0.3cm} \sum_{j=1}^J p_{m,j} \Pr(\ell_m^{t-1}=\ell | \lambda_j ) \nonumber\\
    \hspace{-0.3cm}&=&\hspace{-0.3cm} \sum_{j=1}^J p_{m,j} \frac{\left( \widetilde{\lambda}_{m,j} \right)^{\ell}}{\ell !} e^{-\widetilde{\lambda}_{m,j}}.
\end{eqnarray*}

On the other hand, for a user in the set $\mathcal{G}_r$, the probability that it departs from the cell during the $t$-th segment is 
$$p_{re} = \int_0^{T_s} \mu_m e^{-\mu_m t} dt = 1-e^{-\mu_m T_s}.$$
Thus, $\widetilde{\ell}_m^{t-1}$ follows
\begin{equation*}
    \Pr(\widetilde{\ell}_m^{t-1} = \ell) = \binom{\widetilde{n}_m^{t-1}}{\ell} (p_{re})^{\widetilde{n}_m^{t-1}-\ell}(1-p_{re})^{\ell},
\end{equation*}
for $\ell=0,1,2,\cdots,\widetilde{n}_m^{t-1}$.

Notice that the gNB switching cannot be too frequent. For a relatively large $T_s$, we have $p_{re} \to 1$ and $\Pr(\widetilde{\ell}_m^{t-1} >0) \to 0$. This implies that the residual users from the set $\mathcal{G}_r$ can be omitted. As a result, we arrive at \eqref{eqB1}. And the average $\widetilde n_m^t$ is $\mathbb{E} [\widetilde n_m^t] = \sum_{j=1}^J p_{m,j} \mathbb{E} (\widetilde n_m^t | \lambda_j ) = \sum_{j=1}^J p_{m,j} \widetilde{\lambda}_{m,j} $.



\section{Proof of the non-decreasing monotonicity of $h(\widetilde{n})$ in $\Gamma^L$ and $\Gamma^U$}\label{sec:AppB}
Let us write $h(\widetilde{n})$ as a function of $\Gamma^L$ and $\Gamma^U$:
\begin{eqnarray*}\label{eqF5}
    h(\widetilde{n},\Gamma^L,\Gamma^U)=
    \begin{cases}
        0,& \hspace{-0.3cm} \widetilde{n} \le \Gamma^L; \\
        \Delta_1 (\widetilde{n}),& \hspace{-0.3cm} \widetilde{n} > \Gamma^U; \\
        \Delta_2 (\widetilde{n}) + H^\prime(\Gamma^L,\Gamma^U), & \hspace{-0.3cm} \Gamma^L < \widetilde{n} \le \Gamma^U ,
    \end{cases}
\end{eqnarray*}
where $H^\prime(\Gamma^L,\Gamma^U) \triangleq \Sigma_1 - \Sigma_0 - \epsilon$ is independent of $\widetilde{n}$.

By treating $\widetilde{n}$ as a positive real value, we have
\begin{eqnarray*}\label{eqF6}
    \begin{cases}
        \Delta_2 (\widetilde{n}) + H^\prime(\Gamma^L,\Gamma^U) = 0,&\widetilde{n}=\Gamma^L,\\
        \Delta_2 (\widetilde{n}) + H^\prime(\Gamma^L,\Gamma^U) = \Delta_1 (\widetilde{n}),&\widetilde{n}=\Gamma^U,
    \end{cases}
\end{eqnarray*}
i.e.,
\begin{eqnarray}\label{eqF7}
    \begin{cases}
        \Delta_2 (\Gamma^L) + H^\prime(\Gamma^L,\Gamma^U) = 0,\\
        \Delta_2 (\Gamma^U) + H^\prime(\Gamma^L,\Gamma^U) = \Delta_1 (\Gamma^U).
    \end{cases}
\end{eqnarray}

For any two real values $\Gamma^{L1}<\Gamma^{L2}$, we analyze $h(\widetilde{n},\Gamma^{L1},\Gamma^U)-h(\widetilde{n},\Gamma^{L2},\Gamma^U)$ as follows:
\begin{enumerate}
    \item When $\widetilde{n} \le \Gamma^{L1}$, we have
    $$h(\widetilde{n},\Gamma^{L1},\Gamma^U)\!=\!0,~h(\widetilde{n},\Gamma^{L2},\Gamma^U)\!=\!0,$$
    hence $h(\widetilde{n},\Gamma^{L1},\Gamma^U)-h(\widetilde{n},\Gamma^{L2},\Gamma^U)=0$.
    
    \item When $\Gamma^{L1} < \widetilde{n} \le \Gamma^{L2}$, we have
    $$h(\widetilde{n},\Gamma^{L1},\Gamma^U)=\Delta_2 (\widetilde{n}) + H^\prime(\Gamma^{L1},\Gamma^U) \le 0,$$
    $$h(\widetilde{n},\Gamma^{L2},\Gamma^U)=0,$$
    then $h(\widetilde{n},\Gamma^{L1},\Gamma^U)-h(\widetilde{n},\Gamma^{L2},\Gamma^U)=\Delta_2 (\widetilde{n}) + H^\prime(\Gamma^{L1},\Gamma^U) \le 0$.
    
    \item When $\Gamma^{L2} < \widetilde{n} \le \Gamma^U$, we have
    $$h(\widetilde{n},\Gamma^{L1},\Gamma^U)=\Delta_2 (\widetilde{n}) + H^\prime(\Gamma^{L1},\Gamma^U),$$
    $$h(\widetilde{n},\Gamma^{L2},\Gamma^U)=\Delta_2 (\widetilde{n}) + H^\prime(\Gamma^{L2},\Gamma^U),$$
    then
    \begin{eqnarray*}\label{eqF8}
        && \hspace{-0.5cm} h(\widetilde{n},\Gamma^{L1},\Gamma^U)-h(\widetilde{n},\Gamma^{L2},\Gamma^U) \\
        = && \hspace{-0.5cm} H^\prime(\Gamma^{L1},\Gamma^U)-H^\prime(\Gamma^{L2},\Gamma^U) \\
        \overset{(a)}{=} && \hspace{-0.5cm} -\Delta_2 (\Gamma^{L1}) + \Delta_2 (\Gamma^{L2})         \overset{(b)}{\le} 0,
    \end{eqnarray*}
    where $(a)$ follows from \eqref{eqF7}; $(b)$ follows because $\Delta_2$ is a monotonically non-increasing function and $\Gamma^{L1}<\Gamma^{L2}$.
    
    \item When $\widetilde{n} > \Gamma^U$, we have
    $$h(\widetilde{n},\Gamma^{L1},\Gamma^U)\!=\!\Delta_1 (\widetilde{n}),~
    h(\widetilde{n},\Gamma^{L2},\Gamma^U)\!=\!\Delta_1 (\widetilde{n}),$$
    hence $h(\widetilde{n},\Gamma^{L1},\Gamma^U)-h(\widetilde{n},\Gamma^{L2},\Gamma^U)=0$.
\end{enumerate}

To summarize, we have $h(\widetilde{n},\Gamma^{L1},\Gamma^U)-h(\widetilde{n},\Gamma^{L2},\Gamma^U) \le 0$ when $\Gamma^{L1}<\Gamma^{L2}$.

Likewise, for $\Gamma^{U1}<\Gamma^{U2}$, we compute $h(\widetilde{n},\Gamma^L,\Gamma^{U1})-h(\widetilde{n},\Gamma^L,\Gamma^{U2})$ as follows.
\begin{enumerate}
    \item When $\widetilde{n} \le \Gamma^L$, we have
    $$h(\widetilde{n},\Gamma^L,\Gamma^{U1})\!=\!0,~h(\widetilde{n},\Gamma^L,\Gamma^{U2})\!=\!0,$$
    hence $h(\widetilde{n},\Gamma^L,\Gamma^{U1})-h(\widetilde{n},\Gamma^L,\Gamma^{U2})=0$.
    
    \item When $\Gamma^L < \widetilde{n} \le \Gamma^{U1}$, we have
    $$h(\widetilde{n},\Gamma^L,\Gamma^{U1})=\Delta_2 (\widetilde{n}) + H^\prime(\Gamma^L,\Gamma^{U1}),$$
    $$h(\widetilde{n},\Gamma^L,\Gamma^{U2})=\Delta_2 (\widetilde{n}) + H^\prime(\Gamma^L,\Gamma^{U2}),$$
    then
    \begin{eqnarray*}\label{eqF10}
        && \hspace{-0.5cm} h(\widetilde{n},\Gamma^L,\Gamma^{U1})-h(\widetilde{n},\Gamma^L,\Gamma^{U2}) \\
        = && \hspace{-0.5cm} H^\prime(\Gamma^L,\Gamma^{U1})-H^\prime(\Gamma^L,\Gamma^{U2}) \\
        \overset{(a)}{=} && \hspace{-0.5cm} \left[\Delta_1 (\Gamma^{U1})-\Delta_2 (\Gamma^{U1})\right] - \left[\Delta_1 (\Gamma^{U2})-\Delta_2 (\Gamma^{U2})\right] \\
        \overset{(b)}{=} && \hspace{-0.5cm} g^U(\Gamma^{U1})-g^U(\Gamma^{U2}) \overset{(c)}{\le} 0,
    \end{eqnarray*}
    where $(a)$ follows from \eqref{eqF7}; $(b)$ follows because
    \begin{eqnarray*}\label{eqF11}
        \hspace{-0.3cm} \Delta_1 (\Gamma^U) \!-\! \Delta_2 (\Gamma^U) \hspace{-0.3cm}&=&\hspace{-0.3cm} f\left[ {(\Gamma^U \!+\! \overline{\lambda} T_s ) {\mathcal{P}}_e } \right] - \\
        && \hspace{-0.3cm} f\left[ {{\mathcal{P}}_\text{static} \!+\! {\mathcal{P}}_\text{switch} \!+\! (\Gamma^U \!+\! \overline{\lambda} T_s) {\mathcal{P}}_d } \right] \\
        &=&\hspace{-0.3cm} g^U(\Gamma^U);
    \end{eqnarray*}
    and $(c)$ follows because $g^U(\Gamma^U)$ is a monotonically increasing function and $\Gamma^{U1}<\Gamma^{U2}$.
    
    \item When $\Gamma^{U1} < \widetilde{n} \le \Gamma^{U2}$, we have
    $$h(\widetilde{n},\Gamma^L,\Gamma^{U1})= \Delta_1 (\widetilde{n}),$$
    $$h(\widetilde{n},\Gamma^L,\Gamma^{U2})= \Delta_2 (\widetilde{n}) + H^\prime(\Gamma^L,\Gamma^{U2}),$$
    then, $h(\widetilde{n},\Gamma^L,\Gamma^{U1})-h(\widetilde{n},\Gamma^L,\Gamma^{U2})=\Delta_1 (\widetilde{n}) - \Delta_2 (\widetilde{n}) - H^\prime(\Gamma^L,\Gamma^{U2}) = g^U(\widetilde{n}) - H^\prime(\Gamma^L,\Gamma^{U2})$. As $g^U(\widetilde{n})$ is a monotonically increasing function and $g^U(\Gamma^{U2}) = H^\prime(\Gamma^L,\Gamma^{U2})$, we have $ g^U(\widetilde{n}) - H^\prime(\Gamma^L,\Gamma^{U2}) \le 0$ when $\widetilde{n} \le \Gamma^{U2}$.
    
    \item When $\widetilde{n} > \Gamma^{U2}$, we have
    $$h(\widetilde{n},\Gamma^L,\Gamma^{U1})\!=\!\Delta_1 (\widetilde{n}),~
    h(\widetilde{n},\Gamma^L,\Gamma^{U2})\!=\!\Delta_1 (\widetilde{n}),$$
    hence $h(\widetilde{n},\Gamma^L,\Gamma^{U1})-h(\widetilde{n},\Gamma^L,\Gamma^{U2})=0$.
\end{enumerate}

To summarize, we have $h(\widetilde{n},\Gamma^L,\Gamma^{U1})-h(\widetilde{n},\Gamma^L,\Gamma^{U2}) \le 0$ when $\Gamma^{U1}<\Gamma^{U2}$.

As a result, $h(\widetilde{n})$ is monotonically non-decreasing in $\Gamma^L$ and $\Gamma^U$.

\section{Proof of the decreasing monotonicity of $\Gamma^U(\epsilon)$}\label{sec:AppC}
\begin{enumerate}
    \item When $\epsilon \ge -g^U(0)$, we have
    \begin{eqnarray*}\label{eqF17}
        g^U\left[\Gamma^U(\epsilon) \right] \hspace{-0.3cm}&=&\hspace{-0.3cm} - \epsilon + H\left[\Gamma^L(\epsilon),\Gamma^U(\epsilon) \right] \\
        &\le&\hspace{-0.3cm} g^U(0) + 0 = g^U(0).
    \end{eqnarray*}
    
    As $g^U(\cdot)$ is a monotonically increasing function, we have $\Gamma^U(\epsilon) \le 0$, hence $\widetilde{n} \ge \Gamma^U(\epsilon), \forall \widetilde{n} \ge 0 $. Therefore, $H=E$ and is a constant, where $E$ is defined in \eqref{eqF13}. In this case, $g^U\left[\Gamma^U(\epsilon) \right] = -\epsilon + E$ is monotonically decreasing in $\epsilon$, hence $\Gamma^U(\epsilon)$ is also monotonically decreasing in $\epsilon$.
    
    \item When $\epsilon < -g^U(0)$, for any two costs $\epsilon_1 < \epsilon_2 < -g^U(0)$, we have
    \begin{eqnarray}\label{eqF18}
        g^U\left[\Gamma^U(\epsilon_1) \right] \hspace{-0.3cm}&=&\hspace{-0.3cm} -\epsilon_1 + H\left[\Gamma^L(\epsilon_1),\Gamma^U(\epsilon_1) \right] \nonumber \\
        &>&\hspace{-0.3cm} -\epsilon_2 + H\left[\Gamma^L(\epsilon_1),\Gamma^U(\epsilon_1) \right] \nonumber \\
        &\triangleq&\hspace{-0.3cm} g^U\left[\Gamma^U(\epsilon_{(1)}) \right] .
    \end{eqnarray}
    In particular, we have $\Gamma^L(\epsilon_{(1)}) < \Gamma^L(\epsilon_1)$  since $g^U(\cdot)$ is a monotonically increasing function. Further, we have
    \begin{eqnarray*}\label{eqF19}
        g^U\left[\Gamma^U(\epsilon_{(1)}) \right] \hspace{-0.3cm}&=&\hspace{-0.3cm} -\epsilon_2 + H\left[\Gamma^L(\epsilon_1),\Gamma^U(\epsilon_1) \right] \\
        &\overset{(a)}{>}&\hspace{-0.3cm} -\epsilon_2 + H\left[\Gamma^L(\epsilon_{(1)}),\Gamma^U(\epsilon_{(1)}) \right] \\
        &\triangleq&\hspace{-0.3cm} g^U\left[\Gamma^U(\epsilon_{(2)}) \right],
    \end{eqnarray*}
    where $(a)$ follows because $H$ is a monotonically no\-n-d\-e\-creasing function in $\Gamma^L$ and $\Gamma^U$, according to Lemma~\ref{lem_H_monotone_nondecreasing}.
    In the above manner, we can construct a sequence $\left\{ {\Gamma^U(\epsilon_{(i)}), ~i=1,2,\cdots} \right\}$, where $\Gamma^U(\epsilon_{(1)})$ is defined in \eqref{eqF18}, and the others follow
    \begin{eqnarray}\label{eqF20}
        g^U\left[ \Gamma^U(\epsilon_{(i)}) \right] \hspace{-0.3cm} &=& \hspace{-0.3cm} -\epsilon_2 + H\left[ \Gamma^L(\epsilon_{(i-1)}),\Gamma^U(\epsilon_{(i-1)}) \right] \nonumber \\
        &\overset{(a)}{>}&\hspace{-0.3cm} -\epsilon_2 + H\left[ \Gamma^L(\epsilon_{(i)} ),\Gamma^U(\epsilon_{(i)}) \right] \nonumber \\
        &=&\hspace{-0.3cm} g^U\left[ \Gamma^U(\epsilon_{(i+1)}) \right],~i= 2,3,\cdots,
    \end{eqnarray}
    where $(a)$ can be proven by mathematical induction. As can be seen, $\left\{ {\Gamma^U(\epsilon_{(i)}),~i=1,2,\cdots} \right\}$ is a monotonically decreasing sequence. In particular, $\forall i \ge 2$,
    \begin{eqnarray*}\label{eqF21}
        g^U\left[ \Gamma^U(\epsilon_{(i)}) \right] \hspace{-0.3cm} &=& \hspace{-0.3cm} -\epsilon_2 + H\left[ \Gamma^L(\epsilon_{(i-1)}),\Gamma^U(\epsilon_{(i-1)}) \right] \\
        &>&\hspace{-0.3cm} g^U(0) + E.
    \end{eqnarray*}
    To summarize, $g^U$ is monotonically non-decreasing and lower bounded by a constant $g^U(0) + E$. This suggests that there exists a lower bound for the sequence  $\left\{ {\Gamma^U(\epsilon_{(i)})} \right\}$. Since $\left\{ {\Gamma^U(\epsilon_{(i)})} \right\}$ monotonically decreases, we can denote its lower bound by $\Gamma^U(\epsilon_{(\infty)})$. In particular, according to \eqref{eqF20}, we can write
    \begin{equation*}\label{eqF22}
        g^U\left[ \Gamma^U(\epsilon_{(\infty)}) \right] = -\epsilon_2 + H\left[ \Gamma^L(\epsilon_{(\infty)}),\Gamma^U(\epsilon_{(\infty)}) \right].
    \end{equation*}
    
    Note that $g^U\left[ \Gamma^U(\epsilon_2) \right] = -\epsilon_2 + H\left[ \Gamma^L(\epsilon_2),\Gamma^U(\epsilon_2) \right]$, 
    we have $\epsilon_{(\infty)} = \epsilon_2$, and
    \begin{equation}\label{eqF23}
        \Gamma^U(\epsilon_2) = \Gamma^U(\epsilon_{(\infty)})
        \overset{(a)}{<} \Gamma^U(\epsilon_{(1)})
        \overset{(b)}{<} \Gamma^U(\epsilon_1) ,
    \end{equation}
    where $(a)$ follows because $\left\{ {\Gamma^U(\epsilon_{(i)})} \right\}$ is a monotonically decreasing sequence; $(b)$ follows from \eqref{eqF18}. As a result, we have $\Gamma^U(\epsilon_2) < \Gamma^U(\epsilon_1)$ for $\forall \epsilon_1 < \epsilon_2 < -\Gamma^U(0)$, i.e., $\Gamma^U(\epsilon)$ is monotonically decreasing with $\epsilon$.
\end{enumerate}

Finally, we conclude that $\Gamma^U(\epsilon)$
is monotonically decreasing in $\epsilon$. 

\section{Proof of Theorem~\ref{thm_lower_bound}}\label{sec:AppD}
When the states are in equilibrium, we can rewrite \eqref{eqA1} as
\begin{eqnarray}\label{eqG1}
    {\overline C}_\pi \hspace{-0.3cm}&=&\hspace{-0.3cm} \lim_{T \to \infty } { \mathbb{E}\left[ \frac{1}{T} \sum_{t = 0}^{T - 1} {\sum_{m = 1}^M {c(s_m^t,a_m^t)}} \right] } \nonumber \\
    &\overset{(a)}{=}&\hspace{-0.3cm} \sum_{m = 1}^M {\mathbb{E} [c(s_m,a_m)]}, 
\end{eqnarray}
where $(a)$ holds because the mean values of the cost will not change with time when in equilibrium, and the superscript $t$ can be omitted. As $\mathcal{P}_\text{switch} \ge 0$, the immediate cost of the $m$-th cell $c(s_m,a_m)$ satisfies the following inequality
\begin{eqnarray*}\label{eqG2}
    c(s_m,a_m) \hspace{-0.3cm} &=& \hspace{-0.3cm} f[ {\mathcal{P}} (s_m,a_m) ] \\
    \hspace{-0.3cm} &\ge& \hspace{-0.3cm} \begin{cases}
        f \left[ {\mathcal{P}}_\text{static} + (\widetilde n_m + \overline{\lambda}_m T_s)  {\mathcal{P}}_d \right], & \text{if}~a_m=1; \\
        f \left[ (\widetilde n_m + \overline{\lambda}_m T_s) {\mathcal{P}}_e \right], & \text{if}~a_m=0.
    \end{cases}\nonumber
\end{eqnarray*}
As shown,  $c(s_m,a_m)$ is only dependent on the ON/OFF state and the number of users. We define a variable $\rho_m(\widetilde n_m)$ to express the probability of turning on the $m$-th gNB given the user number $\widetilde n_m$, and $0 \le\rho_m(\widetilde n_m) \le 1$. Then $\mathbb{E} [c(s_m,a_m)]$ in \eqref{eqG1} can be written as
\begin{eqnarray*}\label{eqG3}
    &&\hspace{-0.2cm} \mathbb{E} [c(s_m,a_m)] \\
    &=&\hspace{-0.2cm} \sum_{\ell=0}^{\infty} \mathbb{E} [c(s_m,a_m) | \widetilde n_m = \ell] \cdot \Pr(\widetilde n_m = \ell) \nonumber \\
    &\ge&\hspace{-0.2cm} \sum_{\ell=0}^{\infty} \left\{ \rho_m(\ell) \cdot f \left[ {\mathcal{P}}_\text{static} + (\ell + \overline{\lambda}_m T_s)  {\mathcal{P}}_d \right] \right. \nonumber \\
    && + \left. [1-\rho_m(\ell)] \cdot f \left[ (\ell + \overline{\lambda}_m T_s)  {\mathcal{P}}_e \right] \right\} \cdot \Pr(\widetilde n_m = \ell) \nonumber \\
    &=&\hspace{-0.2cm} C_m^{(0)} + \sum_{\ell=0}^{\infty} \rho_m(\ell) \cdot \Delta_{2,m} (\ell) \cdot \Pr(\widetilde n_m = \ell) .\nonumber
\end{eqnarray*}
The lower bound of the average cost for the $m$-th cell is given by
\begin{eqnarray}\label{eqG4}
    && L_B (\mathbb{E} [c(s_m,a_m)]) \triangleq \\
    && \min_{\{\rho_m\}} C_m^{(0)} + \sum_{\ell=0}^{\infty} \rho_m(\ell) \cdot \Delta_{2,m} (\ell) \cdot \Pr(\widetilde n_m = \ell). \nonumber
\end{eqnarray}

By setting each $\rho_m(\widetilde n_m)$, the minimum value in \eqref{eqG4} can be achieved. A valid solution is given by
\begin{equation}\label{eqG5}
    \rho_m(\widetilde n_m) = \begin{cases}
        0,& \text{if}~\Delta_{2,m} (\widetilde n_m) \ge 0;\\
        1,& \text{if}~\Delta_{2,m} (\widetilde n_m) < 0.
    \end{cases}
\end{equation}
As $f$ is a non-decreasing function, \eqref{eqG5} can be also written as
\begin{equation*}\label{eqG6}
    \rho_m(\widetilde n_m) = \begin{cases}
        0,& \text{if}~\widetilde n_m \le \frac{{\mathcal{P}}_\text{static}}{{\mathcal{P}}_e-{\mathcal{P}}_d} - \overline{\lambda}_m T_s = \gamma_m^L;\\
        1,& \text{if}~\widetilde n_m > \frac{{\mathcal{P}}_\text{static}}{{\mathcal{P}}_e-{\mathcal{P}}_d} - \overline{\lambda}_m T_s = \gamma_m^L,
    \end{cases}
\end{equation*}
where $\gamma_m^L$ is defined in Theorem~\ref{thm_greedy_structure}. Then the lower bound of the average cost for the $m$-th cell can be calculated as
\begin{equation}\label{eqG7}
    L_B(\mathbb{E} [c(s_m,a_m)]) \!=\! C_m^{(0)} \!+\! \sum_{\ell>\gamma_m^L} \Delta_{2,m} (\ell) \cdot \Pr(\widetilde n_m = \ell). 
\end{equation}

If we ignore the constraint $\sum_{m=1}^M {a_m^t} \le M-K$ in \eqref{eqA2} so that the ON/OFF configurations of gNBs are independent, then the average cost in \eqref{eqG1} follows
\begin{eqnarray*}\label{eqG8}
    {\overline C}_\pi \hspace{-0.3cm}&=&\hspace{-0.3cm} \sum_{m = 1}^M {\mathbb{E} [c(s_m,a_m)]} \\
    &\ge&\hspace{-0.3cm} \sum_{m = 1}^M \sum_{\ell=0}^{\infty} \rho_m(\ell) \cdot \Delta_{2,m} (\ell) \cdot \Pr(\widetilde n_m = \ell) + \sum_{m = 1}^M C_m^{(0)}, \nonumber
\end{eqnarray*}
and the lower bound of the average cost is given by
\begin{eqnarray}\label{eqG9}
    && L_B ({\overline C}_\pi) \triangleq \\
    && \min \sum_{m = 1}^M \sum_{\ell=0}^{\infty} \rho_m(\ell) \cdot \Delta_{2,m} (\ell) \cdot \Pr(\widetilde n_m = \ell) + \sum_{m = 1}^M C_m^{(0)}. \nonumber
\end{eqnarray}

As the ON/OFF configurations of gNBs are independent, $L_B({\overline C}_\pi)$ in \eqref{eqG9} can be calculated by
\begin{equation}\label{eqG10}
    L_B({\overline C}_\pi) = \sum_{m = 1}^M L_B(\mathbb{E} [c(s_m,a_m)]).
\end{equation}

Substituting \eqref{eqG7} into \eqref{eqG10}, we arrive at \eqref{eqG0}.

\section{Proof of Proposition~\ref{prop_uniform_policy_performance}}\label{sec:AppE}
Under the uniform policy, we can rewrite \eqref{eqG1} as
\begin{eqnarray}\label{eqH2}
    \overline{C}_{\text{uniform}} = && \hspace{-0.7cm} \sum_{m = 1}^M {\mathbb{E} [c(s_m,a_m)]} \\
    = && \hspace{-0.7cm} \sum_{m = 1}^M \left[C_m^{(01)} \cdot \Pr(a_m^{t-1}=0,a_m^t=1)\right. \nonumber\\
    && \hspace{-0.1cm} + {C_m^{(11)} \cdot \Pr(a_m^{t-1}=1,a_m^t=1)} \nonumber\\
    && \hspace{-0.1cm} \left. + {C_m^{(0)} \cdot \Pr(a_m^t=0)} \right]. \nonumber
\end{eqnarray}

In each time segment, $K$ gNBs are turned off uniformly at random, we have
\begin{eqnarray}\label{eqH3}
    \Pr(a_m^t=0) = K/M \hspace{-0.3cm}&,&\hspace{-0.3cm} ~\Pr(a_m^t=1) = 1-K/M,\nonumber\\
    \Pr(a_m^{t-1}=0,a_m^t=1) \hspace{-0.3cm}&=&\hspace{-0.3cm} \Pr(a_m^{t-1}=0) \cdot \Pr(a_m^t=1) \nonumber\\
    &=&\hspace{-0.3cm} (1-K/M) \cdot K/M, \nonumber\\
    \Pr(a_m^{t-1}=1,a_m^t=1) \hspace{-0.3cm}&=&\hspace{-0.3cm} \Pr(a_m^{t-1}=1) \cdot \Pr(a_m^t=1) \nonumber\\
    &=&\hspace{-0.3cm} (1-K/M)^2.\nonumber
\end{eqnarray}

Substituting the above equations into \eqref{eqH2}, we arrive at \eqref{eqH1}.

\section{Proof of Proposition~\ref{prop_robin_policy_performance}}\label{sec:AppF}
First, when $K=0$, all the gNBs will be always turned on, hence the long-term average cost is
\begin{equation*}\label{eqH6}
    \overline{C}_{\text{round}} = \sum_{m = 1}^M \mathbb{E} \left\{f\left[ {{\mathcal{P}}_\text{static}\!+\!(\widetilde{n}_m^t\!+\!\overline{\lambda}_m T_s) {\mathcal{P}}_d } \right]\right\}\!=\! \sum_{m = 1}^M C_m^{(11)}.
\end{equation*}

When $K=M$, all the gNBs will be always turned off, so the long-term average cost is
\begin{equation*}\label{eqH7}
    \overline{C}_{\text{round}} \!=\! \sum_{m = 1}^M \! \mathbb{E} \! \left\{\!f\!\left[ (\widetilde{n}_m^t \!+\! \overline{\lambda}_m T_s) {\mathcal{P}}_e \right]\right\} \!=\! \sum_{m = 1}^M \! C_m^{(0)}.
\end{equation*}
    
When $0<K<M$, every $M$ time segments can be viewed as a cycle, and the long-term average cost is equal to the cost in one cycle. Thus, the cost can be calculated as
\begin{eqnarray*}\label{eqH8}
    \hspace{-0.3cm} \overline{C}_{\text{round}} \hspace{-0.3cm}&=&\hspace{-0.3cm} \mathbb{E}\left[ \frac{1}{M} \sum_{t = 0}^{M - 1} {\sum_{m = 1}^M {c(s_m^t,a_m^t)}} \right] \\
    &=&\hspace{-0.3cm} \frac{1}{M} \sum_{m = 1}^M {\sum_{t = 0}^{M - 1} \mathbb{E} [c(s_m^t,a_m^t)]} \nonumber \\
     &=&\hspace{-0.3cm} \frac{1}{M} \sum_{m = 1}^M \left[C_m^{(0)} K \!+\! C_m^{(11)} (M\!-\!K\!-\!1) \!+\! C_m^{(01)}\right]. \nonumber
\end{eqnarray*}

To summarize, we arrive at \eqref{eqH5}.

\bibliographystyle{IEEEtran}
\bibliography{References}

\end{document}